\documentclass[twocolumn]{aastex631}

\newcommand{\thiago}[1]

\newcommand{\sm}{$\rm{M_{\odot}}$}

\newcommand{\be}{\begin{equation}}
\newcommand{\ee}{\end{equation}}
\newcommand{\ben}{\begin{eqnarray}}
\newcommand{\een}{\end{eqnarray}}
\newcommand{\bfg}{\begin{figure}}
\newcommand{\efg}{\end{figure}}

\newcommand{\g}{$G$}

\newcommand{\toi}{TOI-1173~A~$b$}

\shorttitle{An Inflated Super-Neptune in a Wide Binary System}
\shortauthors{Yana Galarza et al.}

\begin{document}

\title{TOI-1173 A $b$: The First Inflated Super-Neptune in a Wide Binary System}

\correspondingauthor{J. Yana Galarza}
\email{jyanagalarza@carnegiescience.edu}

\author[0000-0001-9261-8366]{Jhon Yana Galarza}
\altaffiliation{Carnegie Fellow}
\affiliation{The Observatories of the Carnegie Institution for Science, 813 Santa Barbara Street, Pasadena, CA 91101, USA}

\author[0000-0003-2059-470X]{Thiago Ferreira}
\affiliation{Department of Astronomy, Yale University, 219 Prospect Street, New Haven, CT 06511, USA}

\author[0000-0002-1387-2954]{Diego Lorenzo-Oliveira}
\affiliation{Laborat\'orio Nacional de Astrof\'isica, Rua Estados Unidos 154, 37504-364, Itajubá - MG, Brazil}

\author[0000-0002-4733-4994]{Joshua D. Simon}
\affiliation{The Observatories of the Carnegie Institution for Science, 813 Santa Barbara Street, Pasadena, CA 91101, USA}

\author[0000-0001-6533-6179]{Henrique Reggiani}
\affiliation{Gemini South, Gemini Observatory, NSF's NOIRLab, Casilla 603, La Serena, Chile}

\author[0000-0001-6806-0673]{Anthony L. Piro}
\affiliation{The Observatories of the Carnegie Institution for Science, 813 Santa Barbara Street, Pasadena, CA 91101, USA}

\author[0000-0001-6806-0673]{R. Paul Butler}
\affiliation{Earth and Planets Laboratory, Carnegie Institution for Science, 5241 Broad Branch Road, NW, Washington, DC 20015, USA}

\author[0000-0003-2636-9378]{Yuri Netto}
\affiliation{Center for Radio Astronomy and Astrophysics Mackenzie (CRAAM), Mackenzie Presbyterian University, Rua da Consolação, 896, S\~ao Paulo, Brazil}

\author[0000-0002-1671-8370]{Adriana Valio}
\affiliation{Center for Radio Astronomy and Astrophysics Mackenzie (CRAAM), Mackenzie Presbyterian University, Rua da Consolação, 896, S\~ao Paulo, Brazil}

\author[0000-0002-5741-3047]{David R. Ciardi}
\affil{NASA Exoplanet Science Institute, IPAC, California Institute of Technology, Pasadena, CA 91125 USA}

\author{Boris Safonov}
\affil{Sternberg Astronomical Institute Lomonosov Moscow State University}

\begin{abstract} 

Among Neptunian mass exoplanets ($20-50$ M$_\oplus$), puffy hot Neptunes are extremely rare, and their unique combination of low mass and extended radii implies very low density ($\rho < 0.3$~g~cm$^{-3}$). Over the last decade, only a few puffy planets have been detected and precisely characterized with both transit and radial velocity observations, most notably including WASP-107~$b$, TOI-1420~$b$, and WASP-193 $b$. In this paper, we report the discovery of TOI-1173 A $b$, a low-density ($\rho = 0.195_{-0.017}^{+0.018}$~g~cm$^{-3}$) super-Neptune with $P = 7.06$ days in a nearly circular orbit around the primary G-dwarf star in the wide binary system TOI-1173 A/B. Using radial velocity observations with the MAROON-X and HIRES spectrographs and transit photometry from TESS, we determined a planet mass of $M_{\rm{p}} = 27.4\pm1.7\ M_{\oplus}$ and radius of $R_{\rm{p}} = 9.19\pm0.18\ R_{\oplus}$. TOI-1173 A $b$ is the first puffy Super-Neptune planet detected in a wide binary system (projected separation $\sim 11,400$~AU). We explored several mechanisms to understand the puffy nature of TOI-1173 A $b$, and showed that tidal heating is the most promising explanation. Furthermore, we demonstrate that TOI-1173 A $b$ likely has maintained its orbital stability over time and may have undergone von-Zeipel-Lidov-Kozai migration followed by tidal circularization given its present-day architecture, with important implications for planet migration theory and induced engulfment into the host star. Further investigation of the atmosphere of TOI-1173 A $b$ will shed light on the origin of close-in low-density Neptunian planets in field and binary systems, while spin-orbit analyses may elucidate the dynamical evolution of the system. 

\end{abstract}


\section{Introduction} \label{sec:intro}

The discovery of 51 Peg $b$ \citep{1995Natur.378..355M} revolutionized our understanding of planet formation and evolution due to the surprising result that gas giant planets could be observed relatively close to main-sequence stars. This finding hints not only that exoplanets are formed through core accretion of gaseous protoplanetary disc material \citep{1996Icar..124...62P}, but that it is also possible that planets might originate at a wide distance from their host star and subsequently undergo migration \citep{2003ApJ...598L..55R, 2004ApJ...604..388I}. In a similar aspect, despite exhibiting similar masses to the Solar System gas/ice giant planets, inflated radii planets present an enigma, boasting extended and diffuse atmospheres that defy conventional models for planet formation and evolution (e.g., \citealt{2015ApJ...811...41L, 2014ApJ...792....1L}, and references therein). These planets constitute a category characterized by an uncommon combination of large sizes and exceptionally low densities ($\rho \leq 0.3$ g cm$^{-3}$), aptly named puffy\footnote{Super-puff planets are worlds with core masses ($\le 5 M_{\oplus}$) but with radii comparable to gas-giants ($\ge 4 R_{\oplus}$), leading to extremely low densities ($\leq 0.1$ g cm$^{-3}$).} planets. 

Their proximity to their host stars could be responsible for the heating and expansion of their atmospheres \citep{Batygin:2010ApJ...714L.238B, Pu:2017ApJ...846...47P, Thorngren:2018AJ....155..214T}, which contributes to their observed low density. For objects with equilibrium temperatures exceeding 1000 K, the inflated radius mechanism may be similar to hot Jupiter planets \citep{2021JGRE..12606629F}, i.e., (a) due to thermal contraction of He/H atmospheres \citep{2013ApJ...775..105O, 2014ApJ...792....1L}, (b) atmospheric winds driven by photo-ionisation due to intense UV radiation from the host star \citep{2009ApJ...693...23M}, which can be probed by observing larger transit depths due to atmospheric expansion in the Lyman$-\alpha$ line at $\lambda = 1215.6$~\AA\ signatures in a planet's spectrum \citep{2019A&A...623A.131K, 2023MNRAS.518.4357O}, H$\alpha$ absorption at $\lambda = 6562.8$~\AA\ \citep{2012ApJ...751...86J, 2013ApJ...772..144C, 2017AJ....153..217C}, or the Helium triplet in the near-infrared at $\lambda = 10~833$~\AA\ \citep{Spake:2018Natur.557...68S, 2020A&A...640A..29D, Vissapragada:2020AJ....159..278V, 2022A&A...659A..55O, 2023AJ....165..264B, 2024AJ....167...79K}, and/or (c) photo-chemical hazes in the atmosphere responsible for larger and puffier appearance \citep{2020ApJ...890...93G, 2021ApJ...920..124O}. Nevertheless, these mechanisms may not fully elucidate the large radii observed in the lowest density planets with equilibrium temperatures below 1000 K, which poses an additional challenge in incorporating such phenomena into existing exoplanet formation models.

Puffy planets are extremely rare, with only 5 planets in the intermediate-mass regime ($20 M_{\oplus} \leq M_{p} \leq 50 M_{\oplus}$) with densities below $\rho \leq 0.3$ g cm$^{-3}$ (per The Extrasolar Planets Encyclop{\ae}dia\footnote{\url{https://exoplanet.eu/home/}} as of March 2024) in single stars, and none detected in binary systems. These exoplanets are HATS-8 $b$ \citep{Bayliss:2015AJ....150...49B}, TOI-1420 $b$ \citep{Yoshida:2023AJ....166..181Y}, TOI-2525 $b$ \citep{Trifonov:2023AJ....165..179T}, WASP-107 $b$ \citep{Anderson:2017A&A...604A.110A}, and WASP-193 $b$ \citep{Barkaoui:2023arXiv230708350B}, which were detected by both the radial velocity and transit methods. Planets in this category are intrinsically important in deciphering the mechanisms that lead to runaway gaseous accretion despite their smaller cores, a phenomenon crucial to our understanding of planetary formation and evolution \citep{1993prpl.conf.1061L, 2015A&A...582A.112B}. It is noteworthy that only two objects in this sample (TOI-2525 $b$ and WASP-107 $b$) belong to multi-planet systems, and all orbit stars with effective temperatures above 4400~K.

This paper presents the discovery of a low-density inflated super-Neptune ($M_{\rm{p}} = 27.4\ M_{\oplus}$) found in orbit around an 8.7 Gyr old G-dwarf star, completing each orbit in $\sim7.064$ days. The TOI-1173 system contains two stars, the planet-hosting component TOI-1173 A, and TOI-1173 B, which lacks any detected exoplanets (Yana Galarza et al. 2024, submitted). This system is the second wide binary with a separation greater than 10,000 AU discovered to host planets after HAT-P-4 A \citep[29,500 AU;][]{Mugrauer:2014MNRAS.439.1063M} 

In Section \ref{sec:obs} we describe the spectroscopic and photometric observations of TOI-1173~A with MAROON-X and TESS, respectively. In Section \ref{sec:keplerianfit}, we present a Keplerian model, and in Section \ref{sec:discussion}, we discuss the origin of the close-in low-density nature of TOI-1173 A $b$ plus the relevant timescales for the dynamical evolution of this TOI-1173 A/B system. Lastly, we provide an overview of our findings and discuss future research prospects in Section \ref{sec:conclusions}.

\section{Observations}
\label{sec:obs}
\subsection{TESS Photometry}

\begin{figure}
 \includegraphics[width=\columnwidth]{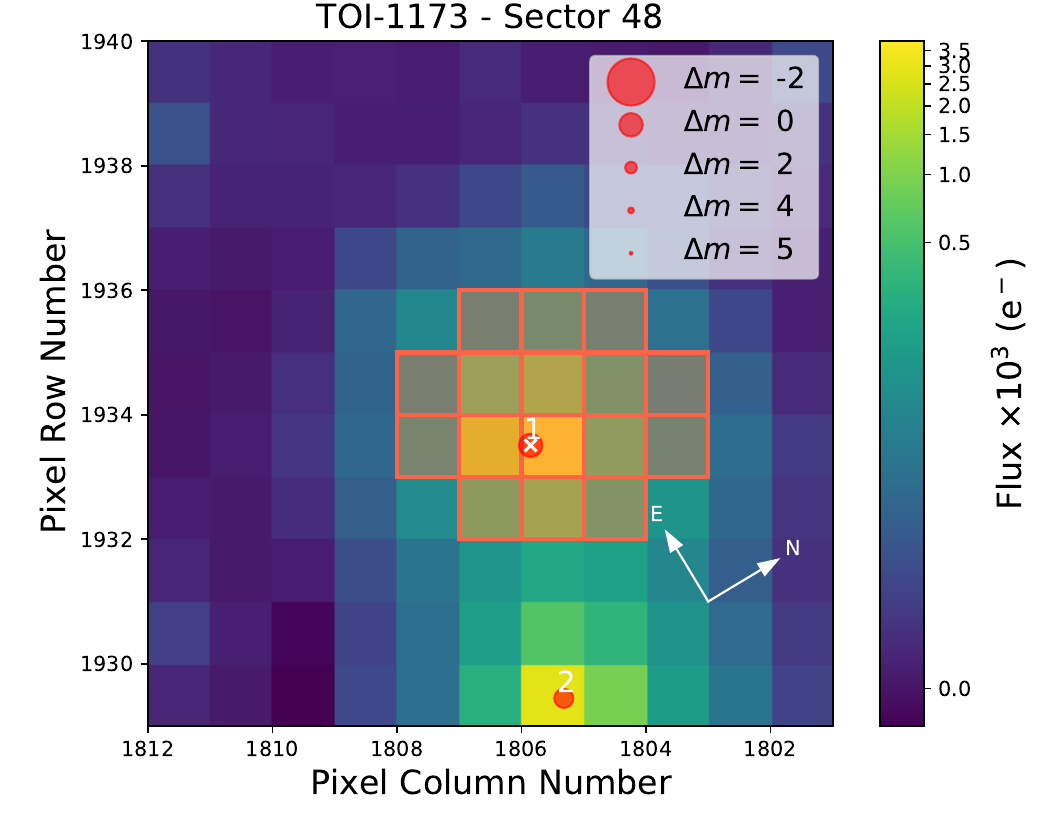}
 \centering
 \caption{Target pixel file of TOI-1173 A (white cross) from the TESS observations in Sector 48. The SPOC pipeline aperture is overplotted with shaded red squares, clearly showing that there is not contamination by the companion star TOI-1173 B (marked with the number 2). The projected separation between the components (numbers 1 and 2) is $\sim$86 arcsec.}
 \label{fig:tpf}
\end{figure}

TOI-1173 A (TIC 232967440) was observed by the Transiting Exoplanet Survey Satellite (TESS; \citealt{2015JATIS...1a4003R}) during Sectors 14, 15, 21, 22, 41, 47, 48, 74, and 75. TESS light curves of TOI-1173 A\footnote{All the TESS data used in this paper can be found in MAST: \dataset[10.17909/dpx3-gv19]{http://dx.doi.org/10.17909/dpx3-gv19}} revealed several periodic transits, leading to the detection and announcement of TOI-1173 A $b$ (TOI-1173.01\footnote{\url{https://exo.mast.stsci.edu/exomast_planet.html?planet=TOI1173.01}}) as a transiting planet candidate by the TESS Science Office (TSO) \citep{Guerrero:2021ApJS..254...39G}. TOI-1173 A is a component of a binary system, and their projected separation is $\sim$86 arcsec \citep{El-Badry:2021MNRAS.tmp..394E}. We used {\sc tpfplotter}\footnote{\url{https://github.com/jlillo/tpfplotter}} \citep{Aller:2020A&A...635A.128A} to check for contamination by the companion star (TOI-1173~B) in the automatically selected aperture. The target pixel file for the pair is displayed in Fig. \ref{fig:tpf}, showing that the components are not blended in TESS. We retrieved the two-minute cadence observations, which were processed by the Science Processing Operations Centre (SPOC; \citealt{2016SPIE.9913E..3EJ}) standard aperture pipeline, which provides systematics-corrected Presearch Data Conditioning (PDC) photometry \citep{Smith:2012PASP..124.1000S, Stumpe:2012PASP..124..985S}, using the {\sc lightkurve} \footnote{\href{https://docs.lightkurve.org/}{https://docs.lightkurve.org/}} software \citep{2018ascl.soft12013L}. 

\begin{figure*}
 \includegraphics[width=2.0\columnwidth]{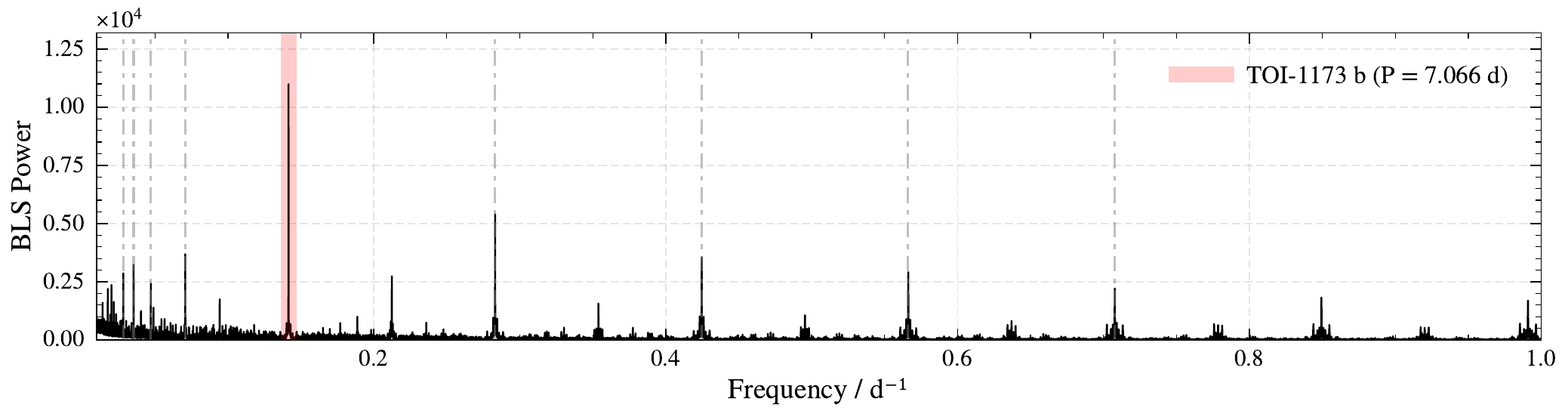}\\
 \includegraphics[width=2.0\columnwidth]{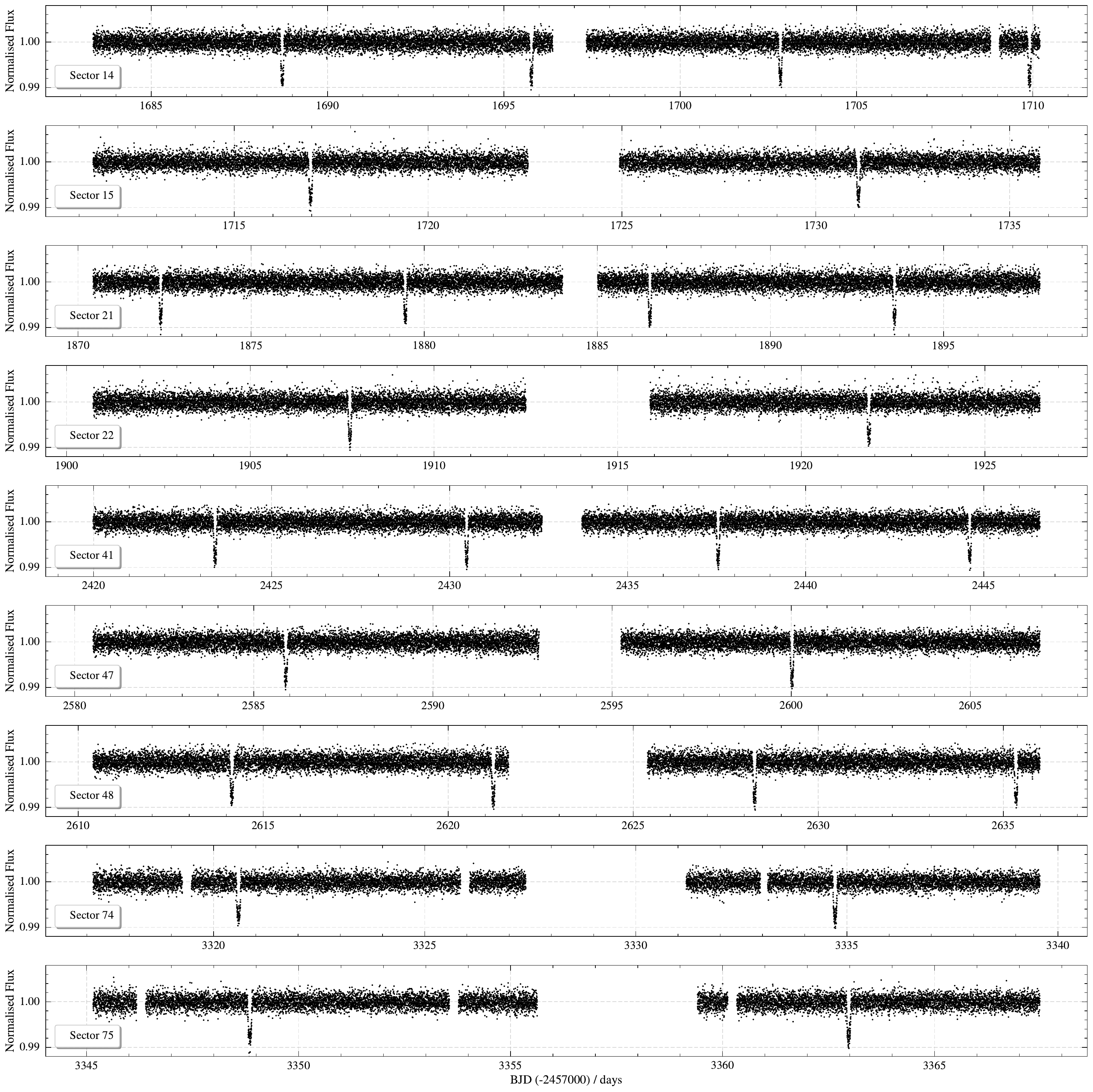}\\
 \centering
 \caption{{\bf Upper panel}: BLS periodogram for TOI-1173 A $b$'s observations. The prominent period of $P = 7.066$ days is indicated as a red strip, along with lower ($P / n$, $n = 1, 2, 3, 4$) and upper ($P\times n$) harmonics in gray. {\bf Bottom panels}: Detrended TESS two-minute cadence PDC light curves. Transit events are visible every 7.066 days.}
 \label{fig:detrend}
\end{figure*}

To remove systematic instrumental trends characterized by a gradual increase/decrease in the apparent stellar flux over time, we applied a de-trending model to the PDC light curve using the 
{\sc w$\bar{o}$tan}\footnote{\url{https://github.com/hippke/wotan}} package \citep{Hippke:2019AJ....158..143H} with a window length of 0.4 and using the Cosine Filtering with Autocorrelation Minimization \citep[CoFiAM,][]{Kipping:2013ApJ...770..101K} algorithm. To minimize the impact of trend removal filter on the transit shape, we mask the transits using the {\sc transit\_mask} feature in {\sc w$\bar{o}$tan}. The PDC detrended light curves of all sectors are depicted in the lower panels of Fig. \ref{fig:detrend}, and we retrieved a prominent modulation period of 7.066 days using a Box-Fitting Least Squares Algorithm (BLS\footnote{\href{https://astrobase.readthedocs.io/en/latest/}{https://astrobase.readthedocs.io/}}; \citealt{2002A&A...391..369K, 2021zndo...4445344B}) in a period grid spanning from 0.25 to 100 days (see upper panel in Figure \ref{fig:detrend}). 

\subsection{MAROON-X and HIRES Radial Velocities}
To confirm the planetary nature of the transit signal, we obtained 10 epochs of spectroscopy of TOI-1173 A with the high-resolution echelle spectrograph MAROON-X \citep{Seifahrt:2018SPIE10702E..6DS, Seifahrt:2022SPIE12184E..1GS} mounted on the 8.1 m Gemini North Telescope of The International Gemini Observatory located in Hawaii (programme ID GN-2022A-Q-227; PI: Yuri Netto). The MAROON-X data were reduced using a custom Python 3 data reduction, which is incorporated into Gemini’s DRAGON platform. The planet-hosting TOI-1173~A was also observed with the High Resolution Echelle Spectrometer (HIRES, \citealp{Vogt:1994SPIE.2198..362V}) at the Keck observatory. The public HIRES spectra were downloaded through Keck Observatory Archive\footnote{\url{https://koa.ipac.caltech.edu/cgi-bin/KOA/nph-KOAlogin}} using the Program IDs: C258\_2020A (PI: Dai), H070\_2020B (PI: Weiss), N062\_2020B (PI: Weiss), C247\_2020B (PI: Dai), U006\_2020B (PI: Robertson), N180\_2021A (PI: Crossfield), N012\_2021A (PI: Howard), U111\_2021A (PI: Batalha), H280\_2021B (PI: Brinkman), U114\_2021B (PI: Dressing), C260\_2021B (PI: Dai), Y166\_2022A (PI: Louden), U072\_2022A (PI: Kane), H258\_2022A (PI: Bottom), C279\_2022A (PI: Zink), and C342\_2022B (PI: Howard). In total, we retrieved 20 spectra with a spectral resolution of $R = \lambda / \Delta \lambda$ = 48,000 and signal-to-noise ratio (SNR) ranging from 100 to 150 on the red chip. The radial velocities were calculated using the iodine cell technique, as described in detail in \citet{Butler:1996PASP..108..500B, Butler:2017AJ....153..208B}. Both the MAROON-X and HIRES radial velocities are listed in Table \ref{tab:MX_RV_data}.

\begin{table}
    \centering
	\caption{Radial velocities for TOI-1173 A collected with the MAROON-X/Gemini and HIRES/Keck spectrographs.}
    \begin{tabular}{lccr}
    \hline 
    \hline
    \multicolumn{4}{c}{\textbf{MAROON-X}} \\
    \hline
    Time (BJD) & $\Delta$RV (m s$^{-1}$) & $\sigma_{\rm RV}$ (m s$^{-1}$) & Channel \\[0.1cm]
    \hline
    2459677.98023 & $-$6.24 & 1.57 & Blue \\
    2459678.95631 & $-$11.74 & 1.54 & Blue\\
    2459680.86707 & $-$6.32 & 1.57 & Blue\\
    2459682.93170 & 8.07 & 1.18 & Blue\\
    2459688.97070 & 1.05 & 1.41 & Blue\\
    2459690.00549 & 5.50 & 1.02 & Blue\\
    2459696.89675 & 6.52 & 1.03 & Blue\\
    2459780.77974 & 0.62 & 1.00 & Blue\\
    2459791.74470 & $-$10.06 & 1.67 & Blue\\
    2459792.74451 & $-$12.01 & 1.27 & Blue\\
    2459677.98023 & $-$10.02 & 2.66 & Red\\
    2459678.95631 & $-$13.25 & 2.60 & Red\\
    2459680.86707 & $-$3.23 & 2.64 & Red\\
    2459682.93170 & 7.36 & 2.04 & Red\\
    2459688.97070 & 1.35 & 2.39 & Red\\
    2459690.00549 & 4.75 & 1.80 & Red\\
    2459696.89675 & 4.09 & 1.80 & Red\\
    2459780.77974 & 3.97 & 1.83 & Red\\
    2459791.74470 & $-$5.79 & 3.00 & Red\\
    2459792.74451 & $-$9.46 & 2.25 & Red\\
    \hline 
    \multicolumn{4}{c}{\textbf{HIRES}} \\
    \hline
    Time (BJD) & $\Delta$RV (m s$^{-1}$) & \multicolumn{2}{c}{$\sigma_{\rm RV}$ (m s$^{-1}$)} \\[0.1cm]
    \hline
    2458885.00826 & 10.14 & \multicolumn{2}{c}{1.70} \\
    2459077.78600 &  0.00 & \multicolumn{2}{c}{1.80} \\
    2459101.73272 & $-$4.98 & \multicolumn{2}{c}{1.82} \\
    2459190.16130 & $-$2.09 & \multicolumn{2}{c}{1.86} \\
    2459215.11314 & $-$2.59 & \multicolumn{2}{c}{1.88} \\
    2459271.95793 & $-$0.59 & \multicolumn{2}{c}{1.95} \\
    2459353.81030 & 11.43 & \multicolumn{2}{c}{1.59} \\
    2459378.85255 & 10.72 & \multicolumn{2}{c}{1.72} \\
    2459435.76640 & 10.10 & \multicolumn{2}{c}{1.80} \\
    2459470.73257 & 11.16 & \multicolumn{2}{c}{1.67} \\
    2459566.15808 & 12.26 & \multicolumn{2}{c}{1.95} \\
    2459593.15571 &  4.02 & \multicolumn{2}{c}{1.73} \\
    2459632.97939 &  9.78 & \multicolumn{2}{c}{1.75} \\
    2459662.02642 &  4.99 & \multicolumn{2}{c}{2.23} \\
    2459690.97466 &  2.52 & \multicolumn{2}{c}{1.92} \\
    2459715.96062 & $-$7.21 & \multicolumn{2}{c}{1.77} \\
    2459739.82355 &  4.05 & \multicolumn{2}{c}{1.76} \\
    2459771.83594 & 16.02 & \multicolumn{2}{c}{1.85} \\
    2459792.76457 & 10.21 & \multicolumn{2}{c}{1.90} \\
    2459826.74058 & $-$8.22 & \multicolumn{2}{c}{1.71} \\
    \hline
    \hline
    \end{tabular}
    \label{tab:MX_RV_data}
\end{table}

We employed the Generalised Lomb-Scargle method (GLS; \citealt{2009A&A...496..577Z}) on a grid ranging from 0.1 to 200 days and detected a highly significant periodic signal at $P_{\rm LS}$ = 7.076 days; which is in line with the period found in the TESS data with a relative difference of $\Delta P = 0.009$ days. A Bayesian Information Criterion test (BIC; \citealt{2014sdmm.book.....I}) revealed a discrepancy of $\Delta{\rm BIC} \approx 400$ between a periodic model and a non-varying model. 

Stellar activity may sometimes simply add noise to the data, but in the worst scenario, it can mimic or masquerade as planetary signals. Therefore, we analysed five stellar activity proxies for TOI-1173 A using the SpEctrum Radial Velocity AnaLyser {\sc SERVAL} pipeline\footnote{\url{https://github.com/mzechmeister/serval}} \citep{2018A&A...609A..12Z}: the differential line width (dLW), chromatic index (CrX) and H$\alpha$ ($\lambda = 656.46$ nm), and the Na D doublet ($\lambda = 589.0$ nm and $589.6$ nm) for the Blue channel, as well as the \ion{Ca}{2} IRT1, IRT2 and IRT3 infrared triplet ($\lambda = 849.8$, $854.2$, and $866.2$ nm) for the Red channel of the MAROON-X spectra. 

\begin{figure*}[t]
    \includegraphics[width = 0.24\linewidth]{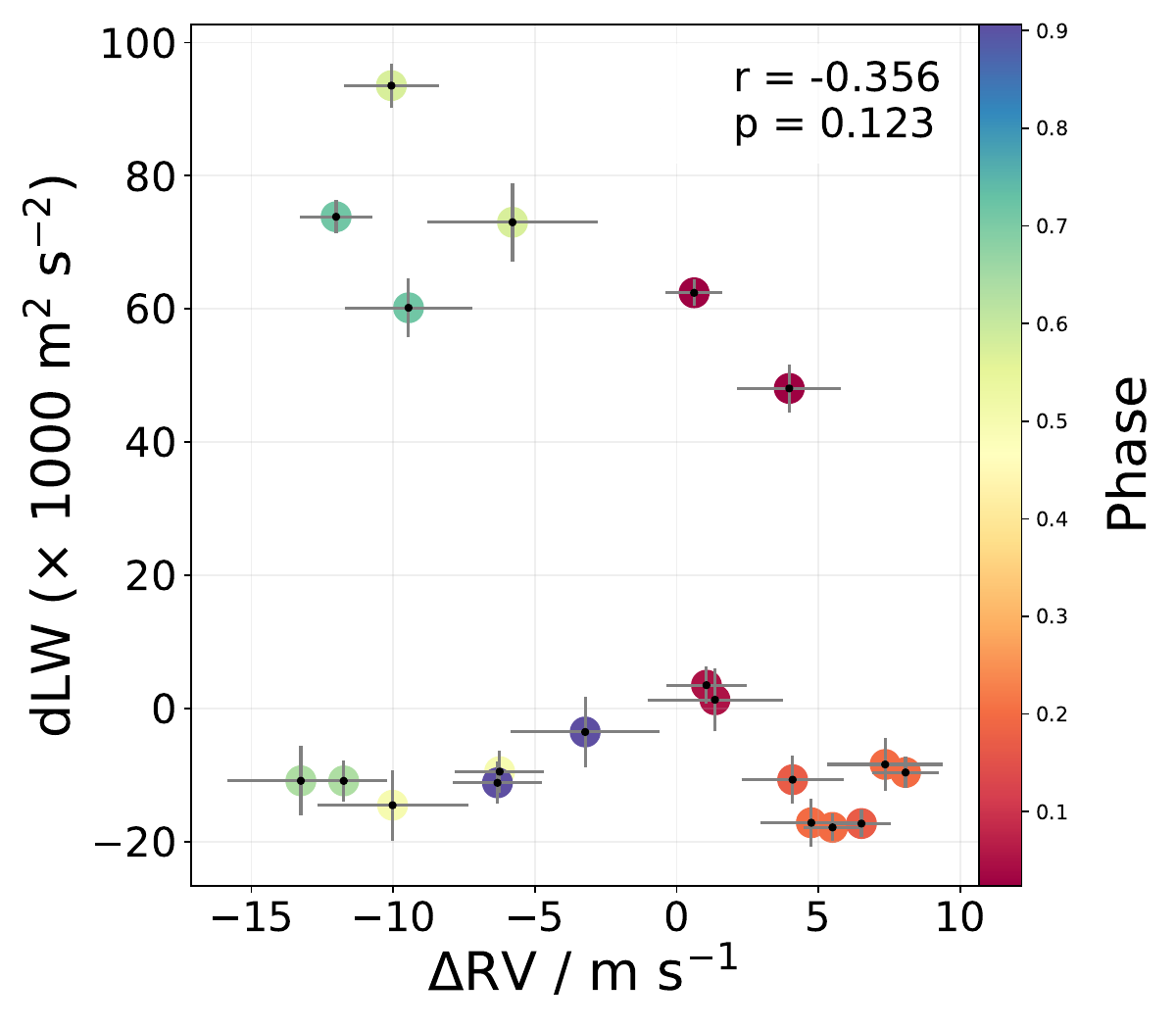}
    \includegraphics[width = 0.24\linewidth]{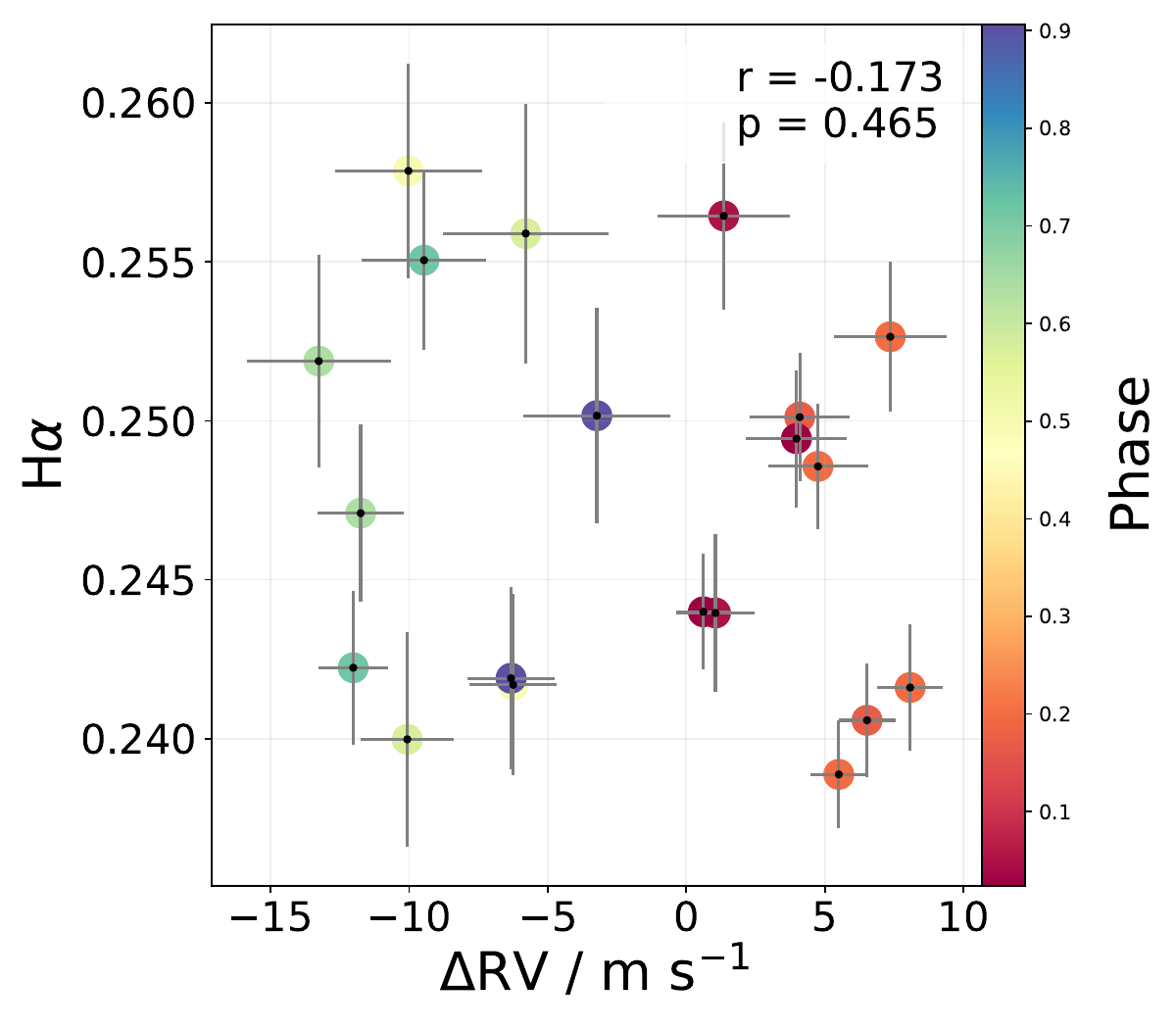}
    \includegraphics[width = 0.24\linewidth]{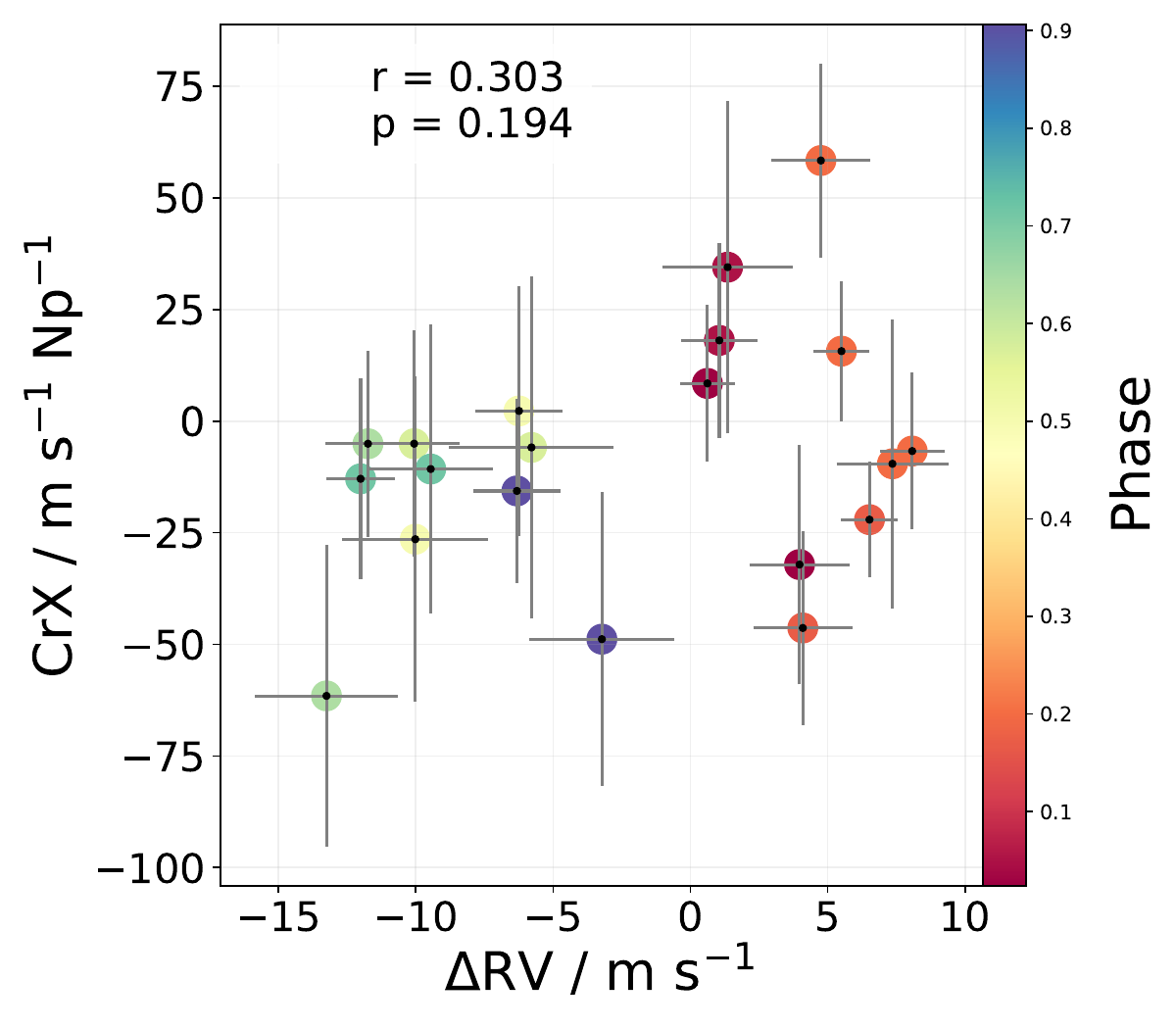}
    \includegraphics[width = 0.24\linewidth]{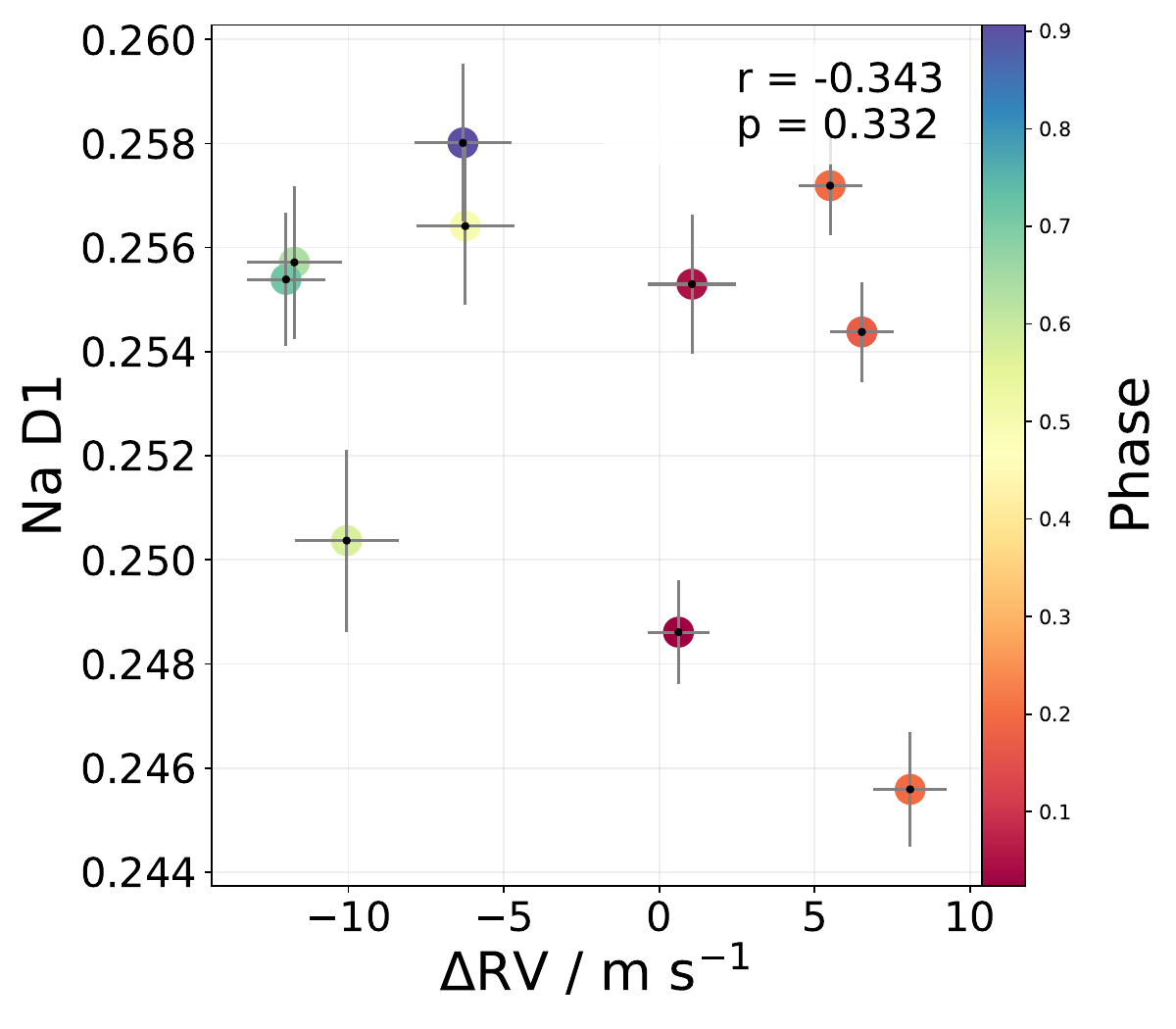}\\
    \includegraphics[width = 0.24\linewidth]{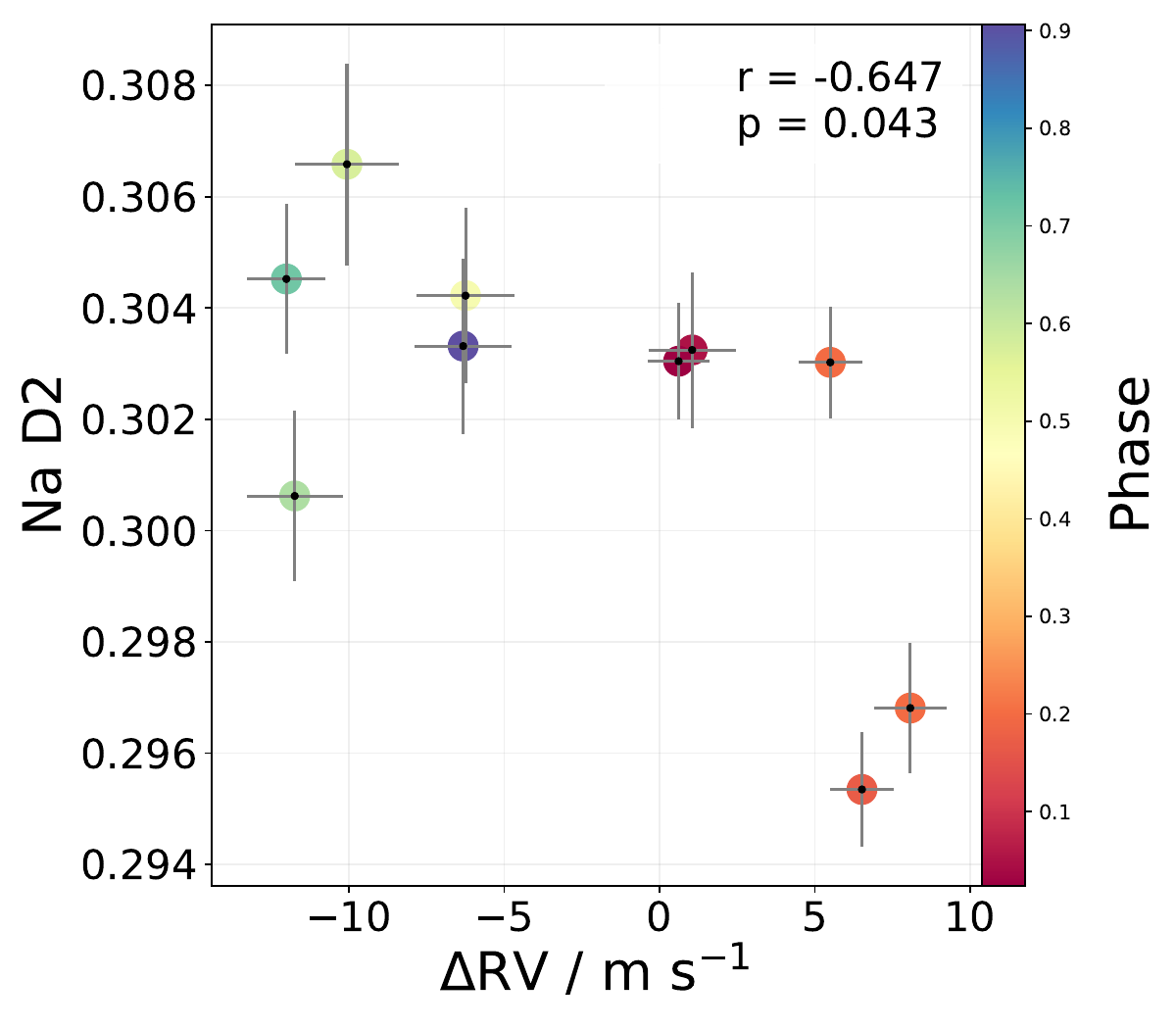}
    \includegraphics[width = 0.24\linewidth]{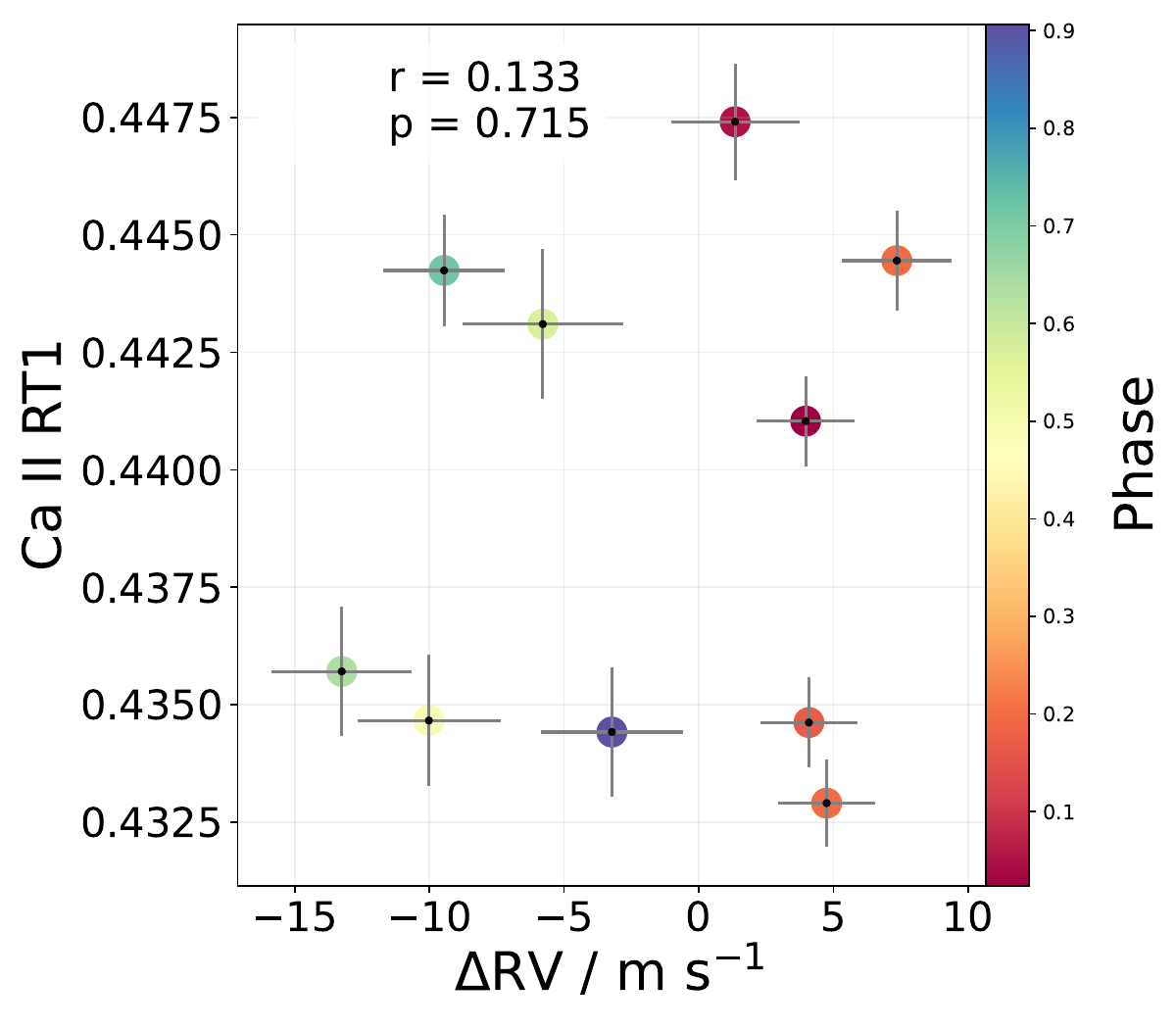}
    \includegraphics[width = 0.24\linewidth]{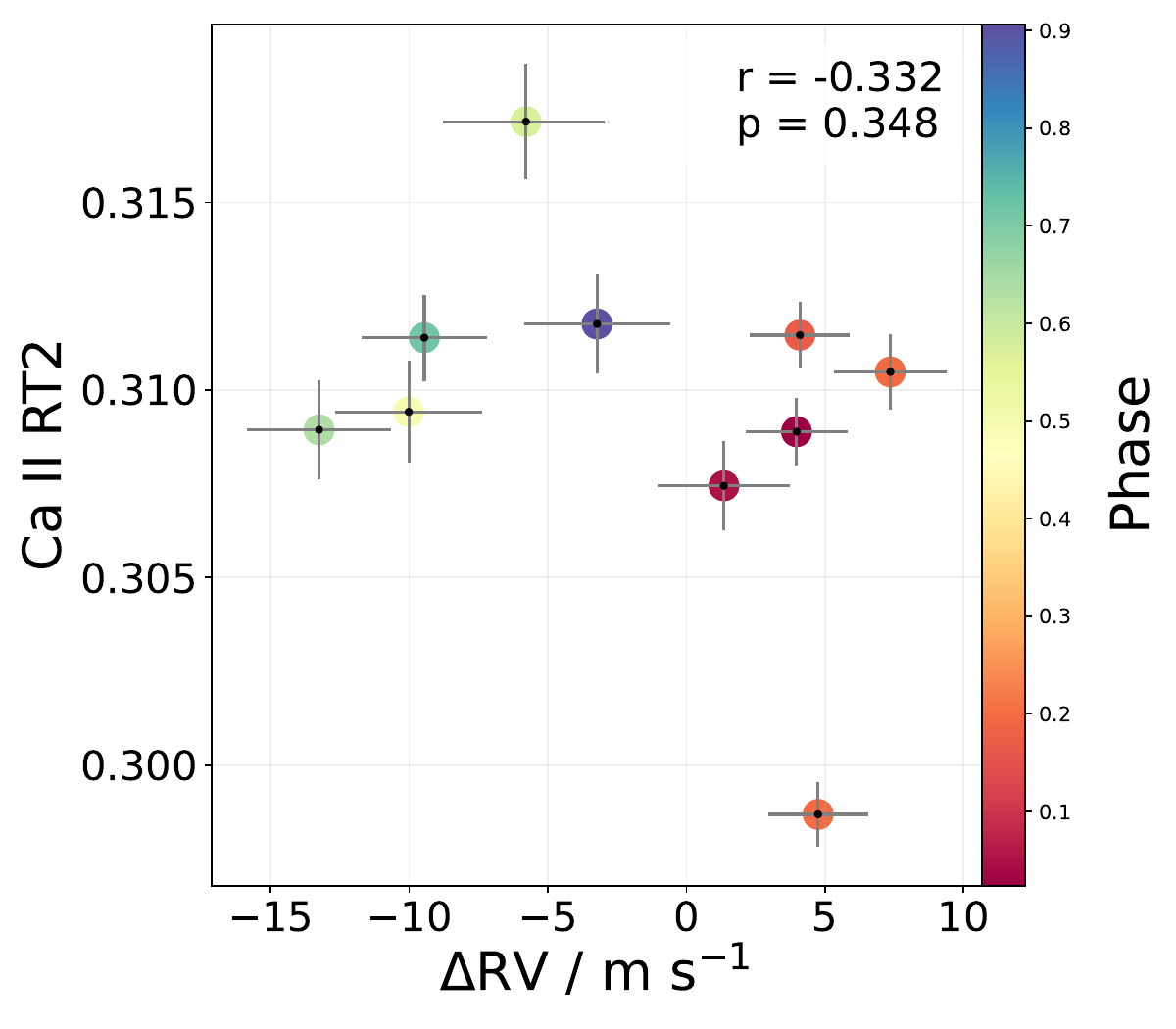}
    \includegraphics[width = 0.24\linewidth]{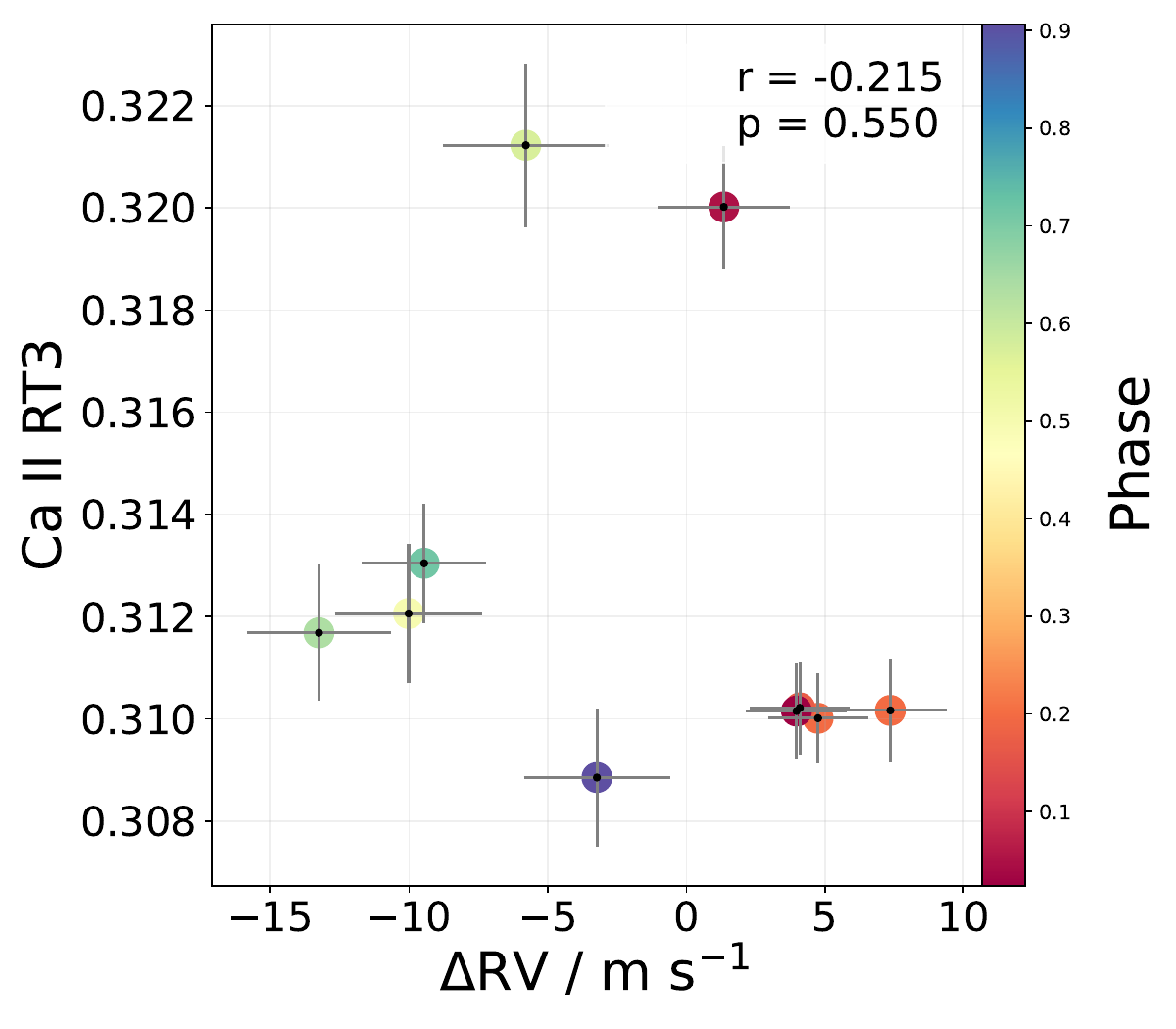}
    \centering
    \caption{Radial velocity and the relationship with TOI-1173 A's activity indicators dLW, CrX, H$\alpha$, plus Na doublet for the Blue channel, and Ca II infrared triplet for the Red channel of MAROON-X spectra. Colors indicate the phase at $P_{\rm LS} = 7.076$ days. The Pearson$-r$ correlation index and the associated $p-$values are indicated at the top of each panel.}
    \label{fig:RVActivity}
\end{figure*}

At the companion's orbital period, we did not observe any indication of significant timing or phase modulations, which could commonly arise if the period were due to stellar activity cycles, such as atmospheric expansions or pulsations (e.g., \citealt{1960mes..book.....S, 1970A&A.....9..245D}). Furthermore, we did not detect statistically meaningful correlations between these magnetic activity indicators and the radial velocity measurements, as evaluated by the Pearson$-r$ correlation index and the null-hypothesis significance testing $p-$value (see Figure \ref{fig:RVActivity}). This analysis strengthens the evidence in favour of a companion nature for the observed radial velocity variations at the retrieved prominent modulation period, and even in the presence of any significant stellar magnetic activity, we do not expect suppression of the companion's signal. 

\subsection{Stellar parameters of the binary system}

TOI-1173 A is the primary component of the binary system TOI-1173 A/B. This pair was first reported as a wide binary system by \citet{El-Badry:2021MNRAS.tmp..394E}. The two stars are separated by approximately $\sim$86 arcsec in the sky, and only the A component harbors an exoplanet. We obtained MAROON-X spectra for both components with high spectral resolution ($R$ = 85,000) and signal-to-noise ratio (SNR $\sim$ 300). We determined the stellar parameters of the pair, meaning effective temperature, surface gravity, metallicity, microturbulence velocity, using the differential approach. We estimated the mass, radius, and age using the isochrone fitting method. The methodology is thoroughly explained in our companion paper  (Yana Galarza et al., 2024, submitted). The results are listed in Table \ref{tab:fundamental_parameters}. Both components are old, and slightly metal-rich stars, with TOI-1173 A being a G9-type star and TOI-1173 B a K1.5-type star.

\begin{table}
    \centering
	\caption{Fundamental parameters of the wide binary system TOI-1173 A/B.} 
    \begin{tabular}{lcr}
    \hline 
	\hline
    {\bf Parameter}                      & {\bf TOI-1173 A}           &  {\bf TOI-1173 B}    \\
    \hline
    Effective temperature (K)            & $5350 \pm 34             $  &  $5047 \pm 34             $    \\
    Surface gravity (dex)                & $4.450 \pm 0.020         $  &  $4.53 \pm 0.030          $    \\
    Metallicity (dex)                    & $0.139 \pm 0.065         $  &  $0.114 \pm 0.069         $    \\
    Microturbulence (km s$^{-1}$)        & $1.11 \pm 0.01           $  &  $1.09 \pm 0.01           $    \\
    Age ($\Gamma$, Gyr)                  & $8.7^{+2.0}_{-1.9}       $  &  $8.6^{+3.1}_{-3.5}        $    \\
    Mass (\sm )                          & $0.911^{+0.028}_{-0.030} $  &  $0.839^{+0.033}_{-0.036} $    \\
    Radius ($R_{\odot}$)                 & $0.934^{+0.011}_{-0.011} $  &  $0.820^{+0.010}_{-0.011} $    \\
    $v \sin i$ (km s$^{-1}$)             & $< 2.0$                     &  $< 2.0$                       \\
    \hline
    \hline
    \end{tabular}
    \label{tab:fundamental_parameters}
\end{table}

\subsection{High-resolution Imaging}

As part of our standard process for validating transiting exoplanets to assess the possible contamination of bound or unbound companions on the derived planetary radii \citep{Ciardi:2015ApJ...805...16C}, we observed TOI-1173 A with high-resolution near-infrared adaptive optics (AO) imaging at Keck Observatory.

\subsubsection{Near-Infrared AO at Keck}

The observations were made with the NIRC2 instrument on Keck-II behind the natural guide star AO system \citep{Wizinowich:2000PASP..112..315W} on 2020-May-28 UT in the standard 3-point dither pattern that is used with NIRC2 to avoid the left lower quadrant of the detector which is typically noisier than the other three quadrants. The dither pattern step size was $3\arcsec$ and was repeated twice, with each dither offset from the previous dither by $0.5\arcsec$. NIRC2 was used in the narrow-angle mode with a full field of view of $\sim10\arcsec$ and a pixel scale of approximately $0.0099442\arcsec$ per pixel. The Keck observations were made in the Br-$\gamma$ filter $(\lambda_o = 2.1686; \Delta\lambda = 0.0326~\mu$m) with an integration time in each filter of 3 seconds for a total of 27 seconds. 

\begin{figure*}
 \includegraphics[width=1.02\columnwidth]{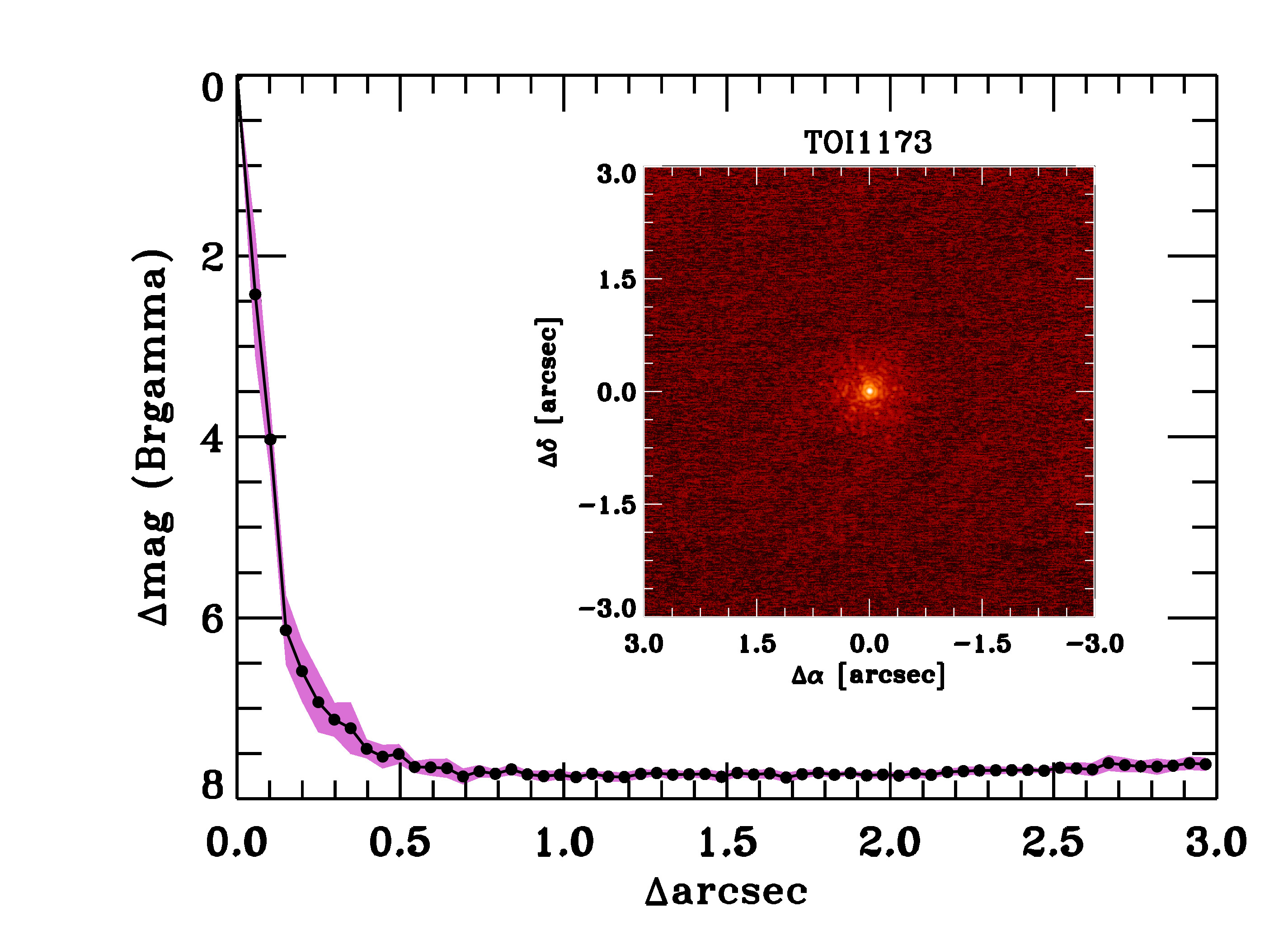}
 \includegraphics[width=\columnwidth]{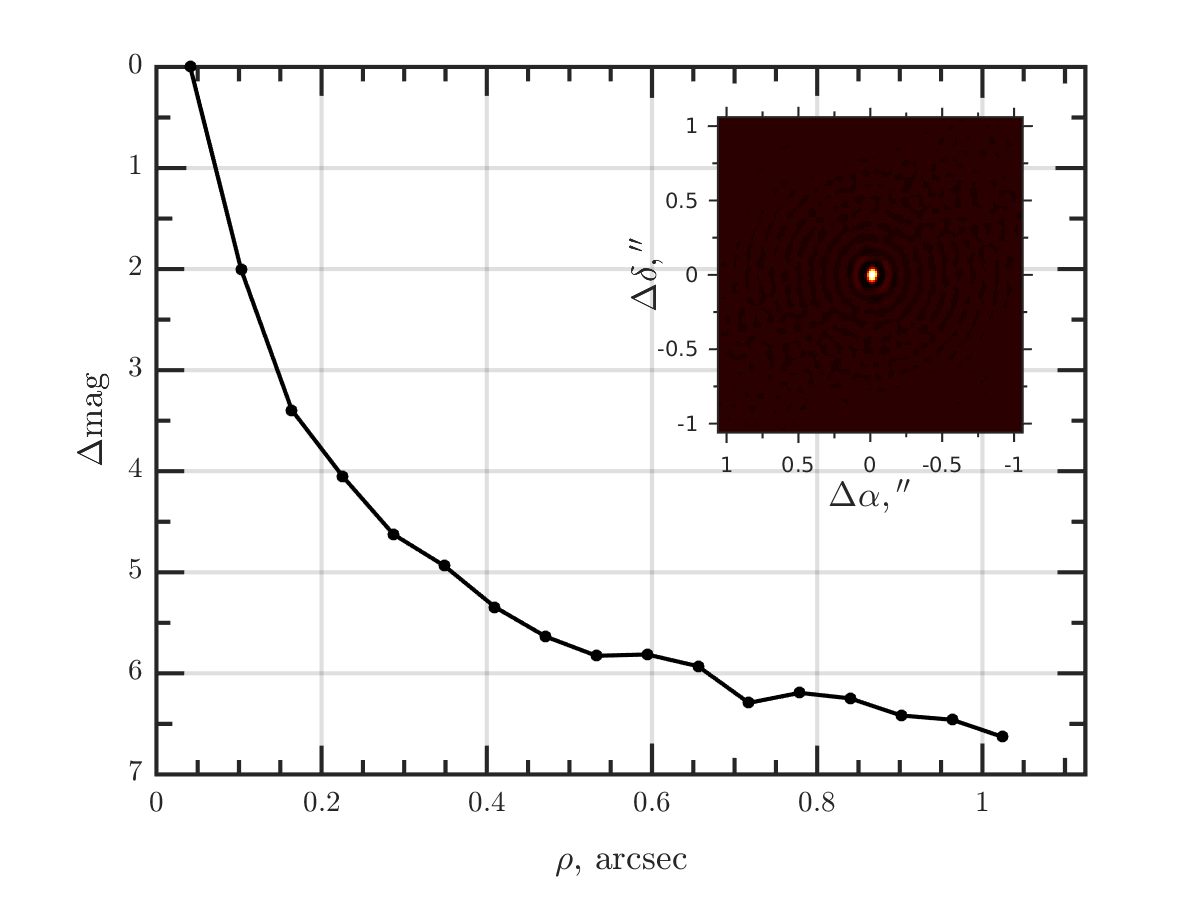}
 \centering
 \caption{\textbf{Left panel:} Companion sensitivity for the near-infrared adaptive optics imaging. The black points represent the 5$\sigma$ limits and are separated in steps of 1 FWHM. The purple represents the azimuthal dispersion (1$\sigma$) of the contrast determinations (see text). The inset image is of the primary target showing no additional close-in companions. \textbf{Right panel:} SAI-2.5m $I_c$-band speckle sensitivity curve for TOI-173 A. The speckle image also does not show evidence of close companions.}
 \label{fig:hri}
\end{figure*}

Flat fields were generated from a median average of dark subtracted dome flats. Sky frames were generated from the median average of the 9 dithered science frames; each science image was then sky-subtracted and flat-fielded. The reduced science frames were combined into a single combined image using an intra-pixel interpolation that conserves flux, shifts the individual dithered frames by the appropriate fractional pixels; the final resolution of the combined dithers was determined from the full-width half-maximum of the point spread function; 0.0494\arcsec. To within the limits of the AO observations, no stellar companions were detected. The final $5\sigma $ limit at each separation was determined from the average of all the determined limits at that separation and the uncertainty on the limit was set by the rms dispersion of the azimuthal slices at a given radial distance (see left panel of Fig.~\ref{fig:hri}).

\subsubsection{Optical Speckle}

We observed TOI-1173 on 2020 November 29 UT with the Speckle Polarimeter \citep{Safonov:2017AstL...43..344S} on the 2.5~m telescope at the Caucasian Observatory of Sternberg Astronomical Institute (SAI) of Lomonosov Moscow State University. The Electron Multiplying CCD Andor iXon 897 was used as a detector. The atmospheric dispersion compensator allowed observation through the wide-band $I_c$ filter. The power spectrum was estimated from 4000 frames with 30 ms exposure times. The detector has a pixel scale of $20.6$ mas pixel$^{-1}$, and the angular resolution was 89 mas. The long-exposure seeing was $1.07\arcsec$. We did not detect any stellar companions brighter than $\Delta I_C=4.3$ and $6.5$ at $\rho=0.25\arcsec$ and $1.0\arcsec$, respectively, where $\rho$ is the separation between the source and the potential companion. The speckle image is depicted in the right panel of Fig. \ref{fig:hri}.

Furthermore, we also found independent analyses in the ExoFOP database based on ground-based time series of TOI-1173 A. These results confirm that the transit is caused by an exoplanet, a conclusion supported by our estimated transit and radial velocity periods.

\section{Keplerian Fit}\label{sec:keplerianfit}

We employed the {\sc allesfitter}\footnote{\href{https://www.allesfitter.com/}{https://www.allesfitter.com/}} software \citep{2021ApJS..254...13G} for a joint transit and radial velocity Keplerian fit of TOI-1173 A $b$ with free parameters including the orbital period ($P$), time of inferior conjunction ($T_0$), Doppler-induced RV semi-amplitude ($K$), the cosine of the orbital inclination ($\cos{i}$), the sum of stellar and companion radii divided by the semi-major axis (($R_p + R_\star) / a$), limb-darkening coefficients $q_1$ and $q_2$ (see \citealt{2013MNRAS.435.2152K} and \citealt{2016MNRAS.457.3573E} for details), plus jitter terms for TESS and both MAROON-X channels. Considering also the possibility of another orbiting companion to TOI-1173 A, or that the modulations were in part due to stellar activity, we set uniform priors for the model barycentre motion $\dot\gamma$ and radial velocity intercept $\gamma$. These parameters are incorporated into the model in the form $\gamma + \dot\gamma\cdot\left(T - T_{0}\right)$, where $T$ represent the time of the observations. Furthermore, we set the orbital eccentricity plus periastron argument coupled as Laplace parameters $\sqrt{e}\cos(\omega)$ and $\sqrt{e}\sin(\omega)$ to avoid Lucy-Sweeney degeneracy \citep{1971AJ.....76..544L}. 

\begin{figure}
    \centering
    \includegraphics[width = \linewidth]{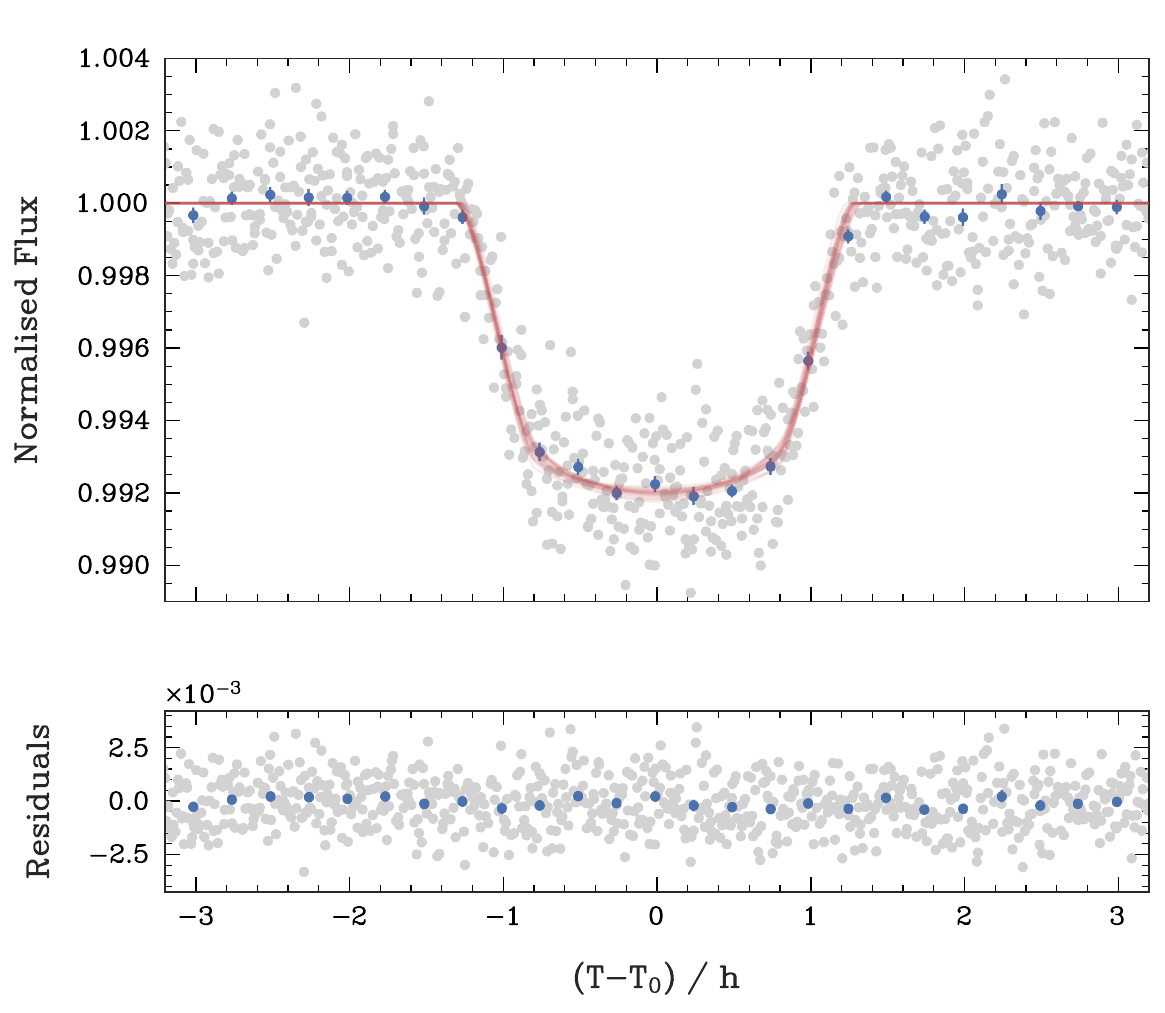}\\
    \includegraphics[width = \linewidth]{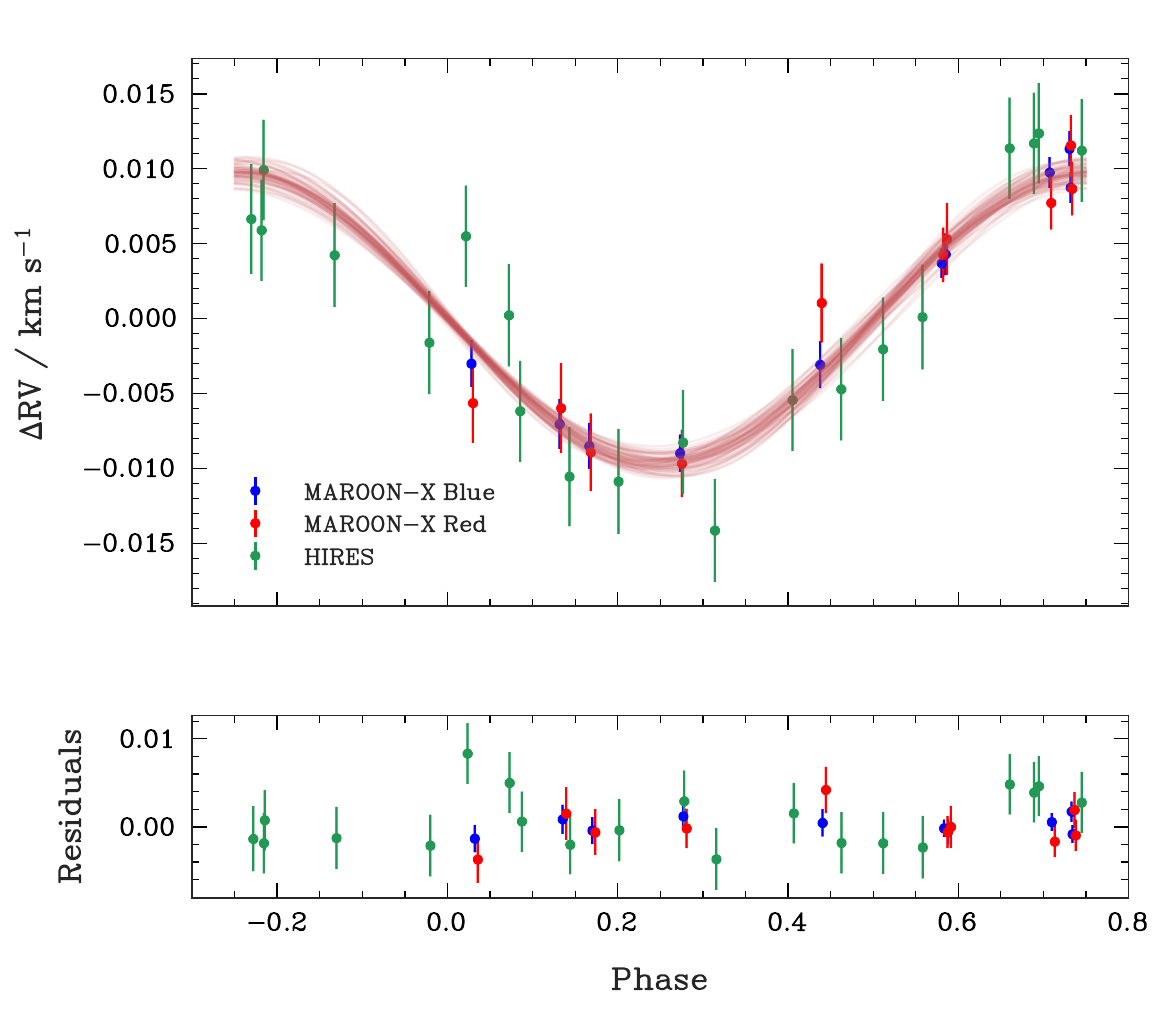}
    \caption{Phase-folded TESS light curve (upper panel) and MAROON-X plus HIRES radial velocities (bottom panel) joint models of TOI-1173A $b$. Red curves show 20 samples drawn from the posterior. The associated residuals are shown below each panel.}
    \label{fig:planet_phase}
\end{figure}

The Markov hyper-parameter space in this Keplerian model was also probed using {\sc DYNESTY} over 500 live points and tolerance of the convergence criterion of $0.01$ (see details in \citealt{2021ApJS..254...13G}). The resulting parameters indicate a planet with mass $27.4\pm1.7$ M$_\oplus$, radius $9.19\pm0.18$ R$_\oplus$ and density $0.195_{-0.017}^{+0.018}$ g cm$^{-3}$ in a nearly circular ($e = 0.023$) orbit around TOI-1173 A every $7.06466$ days (see Table \ref{tab:allesfitter_results} for the full Keplerian parameters). Figure \ref{fig:planet_phase} presents the best-fit 1-companion joint model for TOI-1173 A $b$.

\begin{table*}
    \centering
    \caption{Orbital and physical parameters derived for TOI-1173 A $b$. \\ {\it Note}: $\mathcal{U}(\alpha, \beta)$ stands for a Uniform distribution between $\alpha$ and $\beta$.}
    \begin{tabular}{llll}
    \hline 
	\hline
    {\bf } & {\bf Parameter} & {\bf Values} & {\bf Prior} \\
    \hline
    \multicolumn{3}{l}{{\bf Model Fitted Planetary Parameters}} \\
    Orbital period (d)\dotfill & $P_b$ & $7.06466_{-0.00029}^{+0.00028}$ & $\mathcal{U}$(7.0524, 7.0724) \\ 
    Time of inferior conjunction (d - 2457000)\dotfill & $T_{0;b}$ & $1702.8444_{-0.0023}^{+0.0023}$ & $\mathcal{U}$(1688.662, 1688.762) \\ 
    Doppler-induced RV semi-amplitude (m s$^{-1}$)\dotfill & $K_b$ & $9.67_{-0.46}^{+0.47}$ & $\mathcal{U}$(0, 20) \\ 
    1st Laplace parameter\dotfill & $\sqrt{e_b} \cos{\omega_b}$ & $0.04_{-0.12}^{+0.10}$ & $\mathcal{U}$(-1, 1) \\ 
    2nd Laplace parameter\dotfill & $\sqrt{e_b} \sin{\omega_b}$ & $0.01_{-0.13}^{+0.13}$ & $\mathcal{U}$(-1, 1) \\ 
    Planet-to-star radius ratio\dotfill & $R_b / R_\star$ & $0.0900_{-0.0013}^{+0.0015}$ & $\mathcal{U}$(0, 1) \\ 
    Radii sum divided by semi-major axis\dotfill & $(R_\star + R_b) / a_b$ & $0.0680_{-0.0011}^{+0.0011}$ & $\mathcal{U}$(0, 1) \\ 
    Cosine of the Orbital Inclination\dotfill & $\cos{i_b}$ & $0.0484_{-0.0016}^{+0.0016}$ & $\mathcal{U}$(0, 1) \\ 
    Transformed limb darkening\dotfill & $q_{1; \mathrm{TESS}}$ & $0.28_{-0.11}^{+0.17}$ & $\mathcal{U}$(0, 1) \\ 
    Transformed limb darkening\dotfill & $q_{2; \mathrm{TESS}}$ & $0.36_{-0.27}^{+0.38}$ & $\mathcal{U}$(0, 1) \\ 
    TESS jitter (ppm)\dotfill & $\ln{\sigma_\mathrm{TESS}}$ & $-6.721\pm0.023$ & $\mathcal{U}$(-15, 0) \\ 
    MAROON-X Blue channel jitter (km s$^{-1}$)\dotfill & $\ln{\sigma_\mathrm{jit.; Blue}}$ & $-10.7_{-2.8}^{+2.6}$ & $\mathcal{U}$(-15, 0) \\ 
    MAROON-X Red channel jitter (km s$^{-1}$)\dotfill & $\ln{\sigma_\mathrm{jit.; Red}}$ & $-10.3_{-3.1}^{+2.8}$ & $\mathcal{U}$(-15, 0) \\ 
    HIRES jitter (km s$^{-1}$)\dotfill & $\ln{\sigma_\mathrm{jit.; HIRES}}$ & $-5.88_{-0.24}^{+0.24}$ & $\mathcal{U}$(-15, 0) \\ 
    MAROON-X Blue barycentre motion (km s$^{-1}$ d$^{-1}$)\dotfill & $\gamma_{1; \mathrm{Blue}}$ & $-0.00321\pm0.00060$ & $\mathcal{U}$(-1000, 1000) \\ 
    MAROON-X Red barycentre motion (km s$^{-1}$ d$^{-1}$)\dotfill & $\gamma_{1; \mathrm{Red}}$ & $-0.00448\pm0.00096$ & $\mathcal{U}$(-1000, 1000) \\ 
    HIRES barycentre motion (km s$^{-1}$ d$^{-1}$)\dotfill & $\gamma_{1; \mathrm{HIRES}}$ & $0.0006\pm0.0018$ & $\mathcal{U}$(-1000, 1000) \\ 
    MAROON-X Blue RV slope (km s$^{-1}$ d$^{-2}$)\dotfill & $\dot\gamma_{1; \mathrm{Blue}}$ & $-0.0000\pm0.0011$ & $\mathcal{U}$(-100, 100) \\ 
    MAROON-X Red RV slope (km s$^{-1}$ d$^{-2}$)\dotfill & $\dot\gamma_{1; \mathrm{Red}}$ & $0.0047\pm0.0018$ & $\mathcal{U}$(-100, 100) \\ 
    HIRES RV slope (km s$^{-1}$ d$^{-2}$)\dotfill & $\dot\gamma_{1; \mathrm{HIRES}}$ & $-0.0034\pm0.0026$ & $\mathcal{U}$(-100, 100) \\ 
    \hline
    \multicolumn{3}{l}{{\bf Derived Planet $b$ Parameters}} \\
    Semi-major axis $b$ over host radius\dotfill & $a_\mathrm{b}/R_\star$ & $16.02_{-0.26}^{+0.25}$ & $-$ \\
    Semi-major axis\dotfill & $a_\mathrm{b}$ (AU) & $0.0696\pm0.0014$ & $-$ \\ 
    Mass\dotfill & $M_\mathrm{b}$ ($\mathrm{M_{\oplus}}$) & $27.4\pm1.7$ & $-$ \\ 
    Radius\dotfill & $R_\mathrm{b}$ ($\mathrm{R_{\oplus}}$) & $9.19\pm0.18$ & $-$ \\ 
    Companion density\dotfill & $\rho_\mathrm{b}$ (cgs) & $0.195_{-0.017}^{+0.018}$ & $-$ \\ 
    Eccentricity\dotfill & $e$ & $0.023_{-0.016}^{+0.025}$ & $-$ \\
    Orbital Inclination\dotfill & $i_\mathrm{b}$ (deg) & $87.228_{-0.091}^{+0.087}$ & $-$ \\ 
    Mass ratio (M$_{\rm pl.}/M_{\star}$)\dotfill & $q$ & $0.0000903\pm0.0000048$ & $-$ \\ 
    Impact parameter \dotfill & $b$ & $0.771_{-0.025}^{+0.020}$ & $-$ \\ 
    Total transit duration\dotfill & $T_\mathrm{tot;b}$ (h) & $2.586_{-0.039}^{+0.041}$ & $-$ \\ 
    Companion surface gravity\dotfill & $g_\mathrm{b}$ (cgs) & $314_{-22}^{+23}$ & $-$ \\ 
    Equilibrium temperature $b$\dotfill & $T_\mathrm{eq;b}$ (K) & $868\pm11$ & $-$ \\
    Insolation Flux  \dotfill & S$_{\rm{inc}}$ (S$_{\oplus}$) & $132.0 \pm 10$ & $-$ \\  
    Limb darkening coeff.\dotfill & $u_\mathrm{1; TESS}$ & $0.40_{-0.29}^{+0.29}$ & $-$ \\ 
    Limb darkening coeff.\dotfill & $u_\mathrm{2; TESS}$ & $0.13_{-0.34}^{+0.36}$ & $-$ \\ 
    \hline     
    \hline 
    \end{tabular}
    \label{tab:allesfitter_results}
\end{table*}

\begin{figure*}[t]
\includegraphics[width =\linewidth]{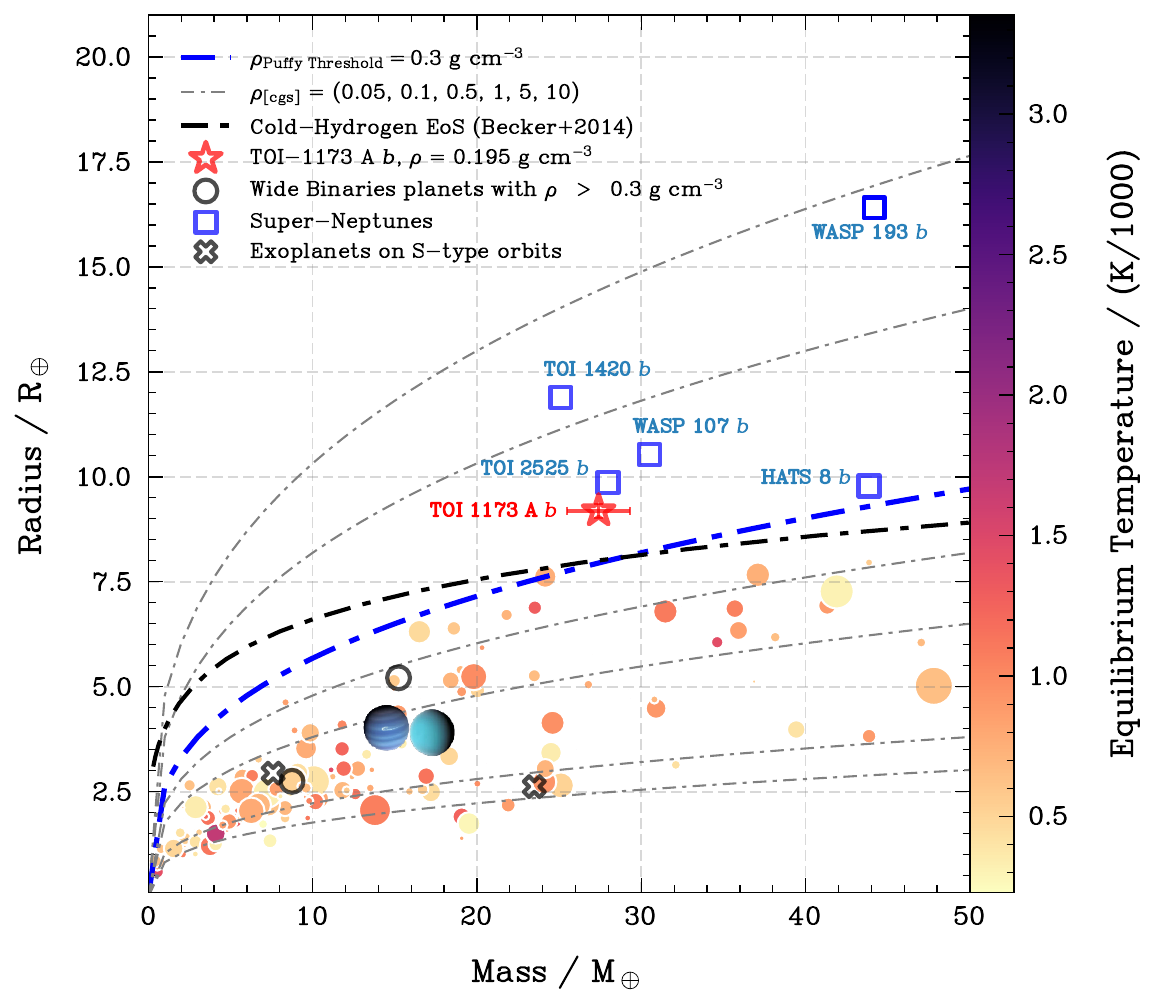}
\centering
\caption{Mass-Radius (M-R) distribution (colored by the planet's equilibrium temperature and sized by the orbital eccentricity) for confirmed exoplanets presented in the Encyclop{\ae}dia of Exoplanetary Systems (\href{https://exoplanet.eu/home/}{https://exoplanet.eu/home/}) as of March 2024. The red star represents TOI-1173 A $b$, and the dashed black line is the theoretical M-R curve for a planet composed of cold Hydrogen according to the H-REOS.3 equation of state from \citet{Becker:2014ApJS..215...21B}, for which TOI-1173 A $b$'s composition seems to be alike. Constant density curves are shown with dot-dashed lines, and the blue one represents the density threshold adopted in this work for puffy planets. In terms of period, semi-major axis, and mass, TOI-1173 A $b$ is similar to TOI-1420 $b$ and WASP-107 $b$. The black open circles represent the planets from the wide binary sample with densities higher than 0.3 g cm$^{-3}$, while black crosses represent planets on S-type orbits (planets that orbit just one star in a binary pair) with masses between $5-35~M_\oplus$.}
\label{fig:MassRadius}
\end{figure*}

\section{Discussion}
\label{sec:discussion}
\subsection{TOI-1173 A $b$: A Low-density Super Neptune}\label{sec:aaa}
In Figure \ref{fig:MassRadius}, we present the discovery of \toi\ in context on a mass-radius diagram of other super-Neptune planets ($20 M_{\oplus} \leq M_{p} \leq 50 M_{\oplus}$) from the Encyclop{\ae}dia of Exoplanetary System as of March 2024. The near neighbours to TOI-1173 A $b$ are TOI-2525 $b$ \citep{Trifonov:2023AJ....165..179T}, WASP-107 $b$ \citep{Piaulet:2021AJ....161...70P} and TOI-1420 $b$ \citep{Yoshida:2023AJ....166..181Y}. The latter two are important targets for studies of planetary atmospheres. Both exoplanets share similar periods, masses, eccentricities and semi-major axes with TOI-1173 A $b$, but have larger radii, which leaves TOI-1173 A $b$ somewhat denser than WASP-107 $b$ and TOI-1420 $b$. Nonetheless, this comparison is relative to single stars. TOI-2525 $b$ is an inflated exoplanet orbiting a K-dwarf star. However, it differs from TOI-1173 A $b$ in its longer period (23 days), higher eccentricity (0.17), and larger semi-major axis (0.15 au). HATS-8 $b$ \citep{Bayliss:2015AJ....150...49B} and WASP-193 $b$ \citep{Barkaoui:2023arXiv230708350B} are also puffy Super-Neptune exoplanets; however, their equilibrium temperatures exceed 1000 K. Remarkably, WASP-193 $b$ has the lowest density among all Super-Neptunes. 
This restricted sample of puffy Super-Neptunes with precisely measured masses and radii makes TOI-1173 A $b$ a valuable addition to this population. 

%Additionally, TOI-1173 A $b$ seems to follow the Cold-Hydrogen Equation-Of-State (EOS) of \citet{Becker:2014ApJS..215...21B} in the mass-radius relation. However, determining its composition and internal structure presents a challenge due to its inflated radius, which can create the impression that the planet shares a similar composition to Saturn and Jupiter.

\begin{figure}
 \includegraphics[width=\columnwidth]{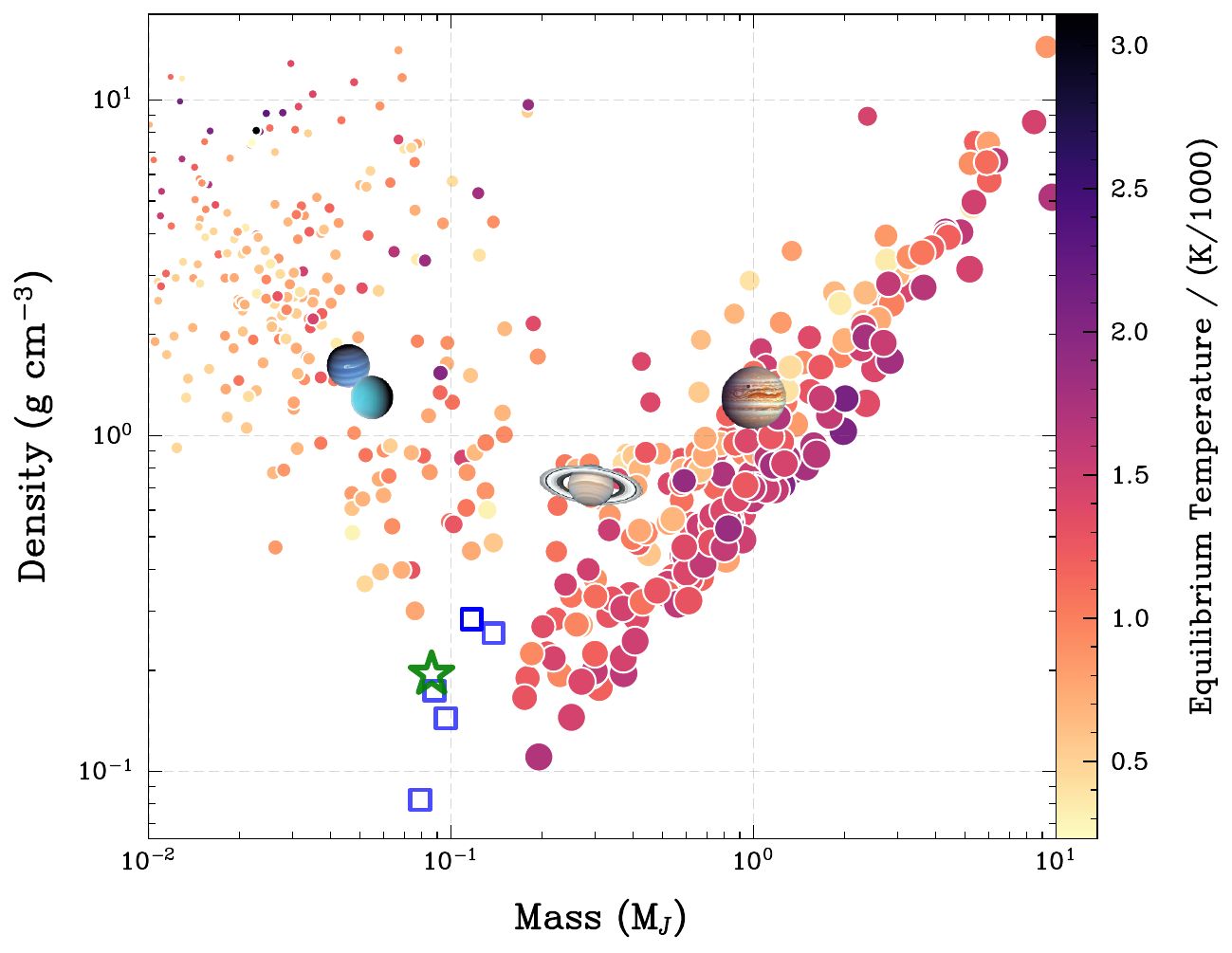}
 \centering
 \caption{Planetary density as a function of the planetary mass of exoplanets obtained from the Encyclop{\ae}dia of Exoplanetary Systems as of March 2024. The characteristic ``V'' shape, and the size of the points, which scale with the planetary radius, denote two regimes of planets. The circles are colored according to their equilibrium temperature. TOI-1173 A $b$ (green star) and the other puffy Super-Neptunes (open blue squares) lie in the transition region.}
 \label{fig:transition}
\end{figure}

Figure \ref{fig:transition} is limited to the giant planets in the range of 0.01-10 $M_{\rm{Jup}}$ and 0.1-1.33 $R_{\rm{Jup}}$. From this figure, we observe that TOI-1173 A $b$, along with other Super-Neptunes, lies in a transition region. On one side are the low-mass, non-degenerate planets, where the bulk density decreases with increasing mass. On the other side are the high-mass, partially degenerate gas giants, where the bulk density increases with increasing mass. Exoplanets in this region are extremely important, as they allow us to explore the properties of planets that have not undergone run-away gas accretion.

When compared with planets in wide binary systems, TOI-1173 A $b$ is the only low-density super Neptune reported to date. The fundamental parameters of this sample of planet-hosting wide binaries were estimated using high resolution ($R > 50,000$) and high signal-to-noise ratio (SNR $> 200$) spectroscopy. They represent the best-characterized sample at present, for which precise masses and radii have been obtained through both radial velocity and transit observations. These exoplanets are: WASP-3 A $b$ \citep{Bonomo:2017A&A...602A.107B, Behmard:2023MNRAS.521.2969B}, HAT-P-1 B $b$ \citep{Turner:2016MNRAS.459..789T, Fan_Liu:2014MNRAS.442L..51L}, WASP-94 A $b$ \citep{Neveu-VanMalle:2014A&A...572A..49N, Teske:2016ApJ...819...19T}, HD 80606 $b$ \citep{Bonomo:2017A&A...602A.107B, Behmard:2023MNRAS.521.2969B}, WASP-160 B $b$ \citep{Lendl:2019MNRAS.482..301L, Emi_Jofre:2021AJ....162..291J}, WASP-127 A $b$ \citep{Seidel:2020A&A...643A..45S, Behmard:2023MNRAS.521.2969B}, HAT-P-4 $b$ \citep{Bonomo:2017A&A...602A.107B, Saffe:2017A&A...604L...4S}, HD 202772 A $b$ \citep{Wang:2019AJ....157...51W, Behmard:2023MNRAS.521.2969B}, Kepler-25 B $b$ and $c$ \citep{Mills:2019AJ....157..145M, Behmard:2023MNRAS.521.2969B}, KELT-2 A $b$ \citep{Stassun:2017AJ....153..136S, Behmard:2023MNRAS.521.2969B},  XO-2N $b$ \citep{Bonomo:2017A&A...602A.107B, Teske:2015ApJ...801L..10T}, WASP-173 $b$ \citep{Labadie-Bartz:2019ApJS..240...13L, Behmard:2023MNRAS.521.2969B}, WASP-180 A $b$ \citep{Temple:2019MNRAS.490.2467T, Behmard:2023MNRAS.521.2969B}, and WASP-64 $b$ \citep{Bonomo:2017A&A...602A.107B, Behmard:2023MNRAS.521.2969B}. Only two of these exoplanets have a mass below 20 $M_{\oplus}$, but with densities $\rho > 0.3$ g cm$^{-3}$, depicted as circles in Fig.~\ref{fig:MassRadius}. Within the super Neptune regime, there exists only one wide binary with two exoplanets, Kepler-25 B $b$ and $c$. However, these exoplanets have substantially higher densities ($\rho \geq 0.5$ g cm$^{-3}$). This makes TOI-1173 A $b$ a potentially unique target for exploring planet formation in wide binary systems.

In the remainder of this section, we investigate different hypotheses to explain the puffy nature of TOI-1173 A $b$.

\subsubsection{Stellar Insolation}
It is well known that the inflated radii of some hot Jupiters ($M \geq 0.3~M_{\rm Jup}$, $P < 10$ days, and $T_{\rm eq} > 1000$ K), are primarily attributed to the high incident stellar flux upon these planets \citep[e.g.,][]{Batygin:2010ApJ...714L.238B, Pu:2017ApJ...846...47P, Thorngren:2018AJ....155..214T}. A fraction of the light from the star penetrates deep into the planetary atmosphere, where it is then absorbed, causing its inflation \citep[see][for more details]{Fortney:2021JGRE..12606629F}. However, as seen in Table \ref{tab:allesfitter_results}, TOI-1173 A $b$ does not share the planetary properties of hot Jupiters, as its insolation and equilibrium temperature are lower. 

To ensure a proper comparison with super-Neptunes precisely characterized by transit and RV observations, we have selected stars with temperatures, metallicities, masses, and radii within $\pm$200 K, $\pm$0.2 dex, $\pm$0.2 \sm, and $\pm$0.2 R$_{\odot}$ of those of TOI-1173 A. Figure \ref{fig:insolation} indicates that the super-Neptune TOI-1173 A $b$ and TOI-1420 $b$ experience higher incident flux from their host star than other similar gas giants, which suggests that the inflated state of TOI-1173 A $b$ could be attributed to stellar insolation. However, assessing whether this insolation is exceptionally high compared to stars with similar properties as TOI-1173 A is challenging due to the restricted sample size.
\subsubsection{Excess of Internal Heat}\label{subsec:excessinternalheat}
Various studies have highlighted the impact of internal heat on a planet's radius, indicating that planets with hotter interiors have larger radii and lower densities than those with cooler interiors of the same composition \citep[e.g.,][]{Bodenheimer:2001ApJ...548..466B, Batygin:2010ApJ...714L.238B, Lopez:2014ApJ...792....1L, Millholland:2019ApJ...886...72M}. This inflation stems from the internal heat remaining from the formation of the 
planet rather than external factors. The age of a planet can influence its internal heat, with younger planets naturally having hotter interiors and consequently larger radii. However, in the case of the $\sim8$ billion-year-old TOI-1173 A/B system, TOI-1173 A $b$ has already cooled \citep{Linder:2019A&A...623A..85L}. Hence, an alternative mechanism must be responsible for inflating its radius.

\subsubsection{Tidal Heating}
Tidal heating results from gravitational forces between celestial bodies, causing internal friction and generating heat within the affected body. \citet{Millholland:2020ApJ...897....7M} demonstrated that tidal interactions at 0.1 AU can yield tidal luminosities up to $L_{\rm{tide}} \sim 10^{29}$ erg s$^{-1}$ (refer to their Figure 1). Applying the tidal luminosity expression from Equation (1) in \citet{Millholland:2020ApJ...897....7M}, we estimate $L_{\rm{tide}} \sim 10^{23}$ erg s$^{-1}$ (assuming zero obliquity and a reduced tidal quality factor $Q' = 10^{5}$ for Neptune like planets \citep{Tittemore:1990Icar...85..394T, Banfield:1992Icar...99..390B}) for TOI-1173 A $b$. This suggests the potential for significant tidal heating on the planet.

\begin{figure}
 \includegraphics[width=\columnwidth]{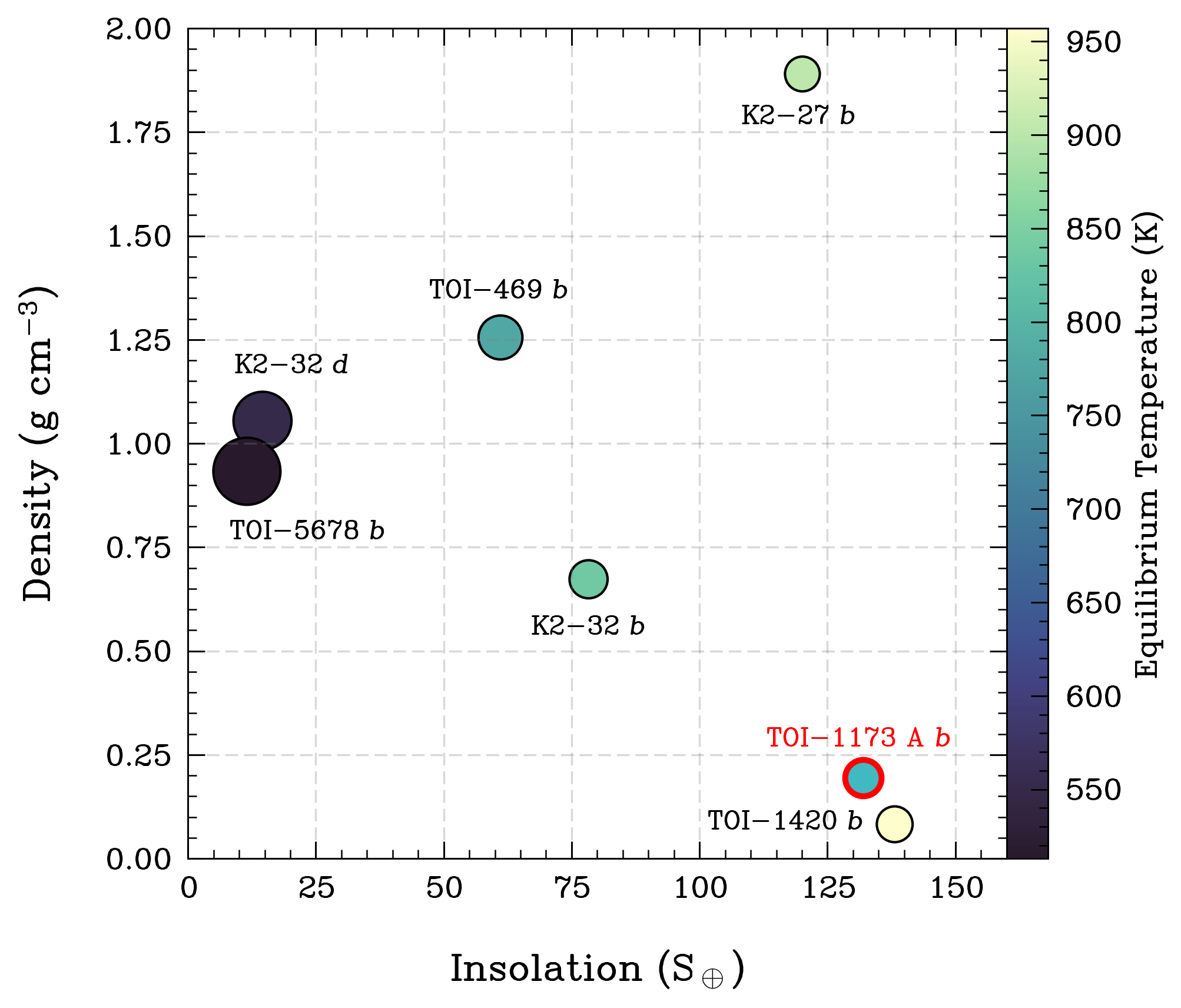}
 \centering
 \caption{Planet density as a function of stellar insolation. TOI-1173 A $b$ is indicated in red. The circles represent the planets with host stars sharing similar physical parameters with TOI-1173 A, colored by their equilibrium temperatures and sized by the planet's semi-major axis.}
 \label{fig:insolation}
\end{figure}

\begin{figure}
 \includegraphics[width=\columnwidth]{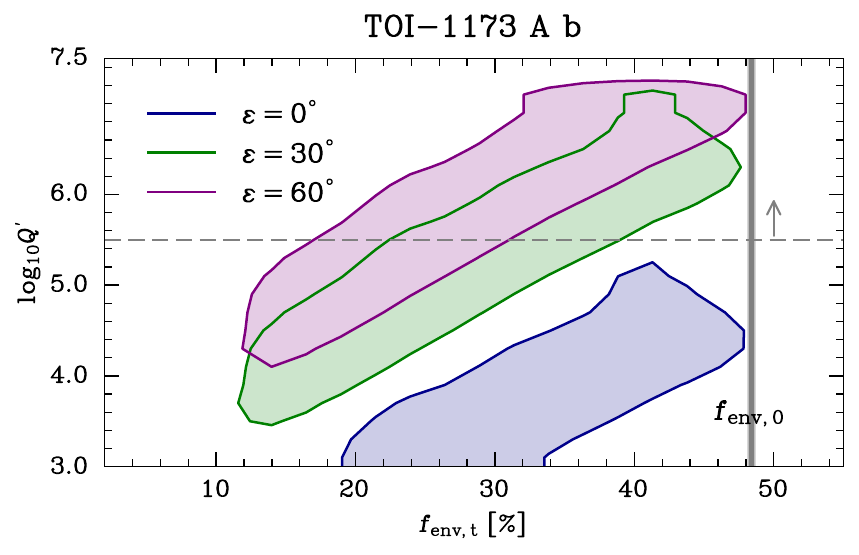}
 \centering
 \caption{Envelope mass fraction estimates with tides ($f_{\rm{env}},t$) and without tides ($f_{\rm{env}},0$) of TOI-1173 A $b$. The horizontal dashed lines represent the lower limit of the reduced tidal quality factor ($\log_{10} Q'$). The mean and standard deviation of $f_{\rm{env}},0$ are represented by the vertical gray bar. Colored regions indicate the $2\sigma$ contours of the posterior distributions of $\log_{10} Q'$ and $f_{\rm{env}},t$ after accounting for tides.}
 \label{fig:envelope}
\end{figure}

To further investigate this mechanism, we used the structural model with tidal heating of \citet{Millholland:2019ApJ...886...72M}. This model incorporates tidal heating effects on planetary structures by considering a two-layer planet model consisting of a heavy element core and an H/He envelope with an MCMC fitting approach. The atmospheric envelope evolution is simulated using the Modules for Experiments in Stellar Astrophysics (MESA) code \citep{Paxton:2011ApJS..192....3P, Paxton:2013ApJS..208....4P, Paxton:2015ApJS..220...15P, Paxton:2018ApJS..234...34P, Paxton:2019ApJS..243...10P, Jermyn2023} within Sub-Saturns\footnote{ \url{https://github.com/smillholland/Sub-Saturns/}} code \citep{Millholland:2020ApJ...897....7M}. Figure \ref{fig:envelope} shows the envelope mass fraction estimates for TOI-1173 A $b$. The colored regions indicate the 2D posterior distributions for $\log_{10} Q'$ and $f_{\rm{env}, t}$ obtained from the fitting that includes tidal inflation. The horizontal dashed lines indicate the lower limit of the reduced tidal quality factor ($\log_{10} Q' \sim 5.5 $). These distributions assume the measured eccentricities and indicate the results for obliquity $\epsilon =$ 0\textdegree, 30\textdegree and 60\textdegree. The mean and 1$\sigma$ range of $f_{\rm{env}},0$ inferred when neglecting tides are shown with the gray line and bar. 

Our findings indicate that, when not considering tidal effects, TOI-1173 A $b$ has an envelope mass fraction $f_{\rm{env}},0$ of $48.4 \% \pm 0.3 \%$. However, incorporating tidal heating into the model led to a significant decrease in the estimated envelope mass fraction, ranging from $f_{\rm{env}},t$ = 18\%--40\% for $\epsilon =$ 30\textdegree--60\textdegree. This reduction in the envelope mass fraction is attributed to the impact of tidal inflation resulting from eccentricity and obliquity tides. For $\epsilon =$ 0\textdegree, it seems that tidal heating has no impact on the planet, as the 2D posterior distribution is below the lower limit of the estimated reduced tidal quality factor $\log_{10} Q' = 5.5$ (horizontal dashed lines in Fig. \ref{fig:envelope}).

\subsubsection{Planetary Rings}
\citet{Piro:2020AJ....159..131P} suggested that the inflated radii of low-density puffy planets could be a result of planetary rings, which would cause deeper-than-expected transits. This hypothesis is contingent upon the rings being at an oblique angle to the planet's orbital plane. Additionally, if the planet is tidally locked, the impact on the overall transit depth would be negligible. Using the Equation (11) from \citet{Piro:2020AJ....159..131P}, we determined a synchronous timescale of 0.061 Myr for TOI-1173 A $b$, confirming its tidal locking. If TOI-1173 A $b$ has rings, they would lie in its orbital plane and thus be observed edge-on from our vantage point, thus not explaining the apparent inflated radius. Moreover, because of the planet's proximity to its host star, the rings would be susceptible to dissipating or collapsing under various gravitational forces and interactions \citep{Ohta:2009ApJ...690....1O}. 

\subsubsection{Mass Loss by Atmospheric Escape}

Given the relatively high stellar irradiation of TOI-1173 A $b$ (see Figure \ref{fig:insolation}) and its proximity to its star, it is reasonable to assume that the planet is losing mass. Therefore, the inflated radius may be also attributed to material evaporating and leaving the planet \citep[e.g.,][]{Guo:2013ApJ...766..102G, Ehrenreich:2015Natur.522..459E, Koskinen:2022ApJ...929...52K}. \citet{Fossati:2017A&A...598A..90F} demonstrated that it is possible to use theoretical models of mass loss to infer whether atmospheric escape is significant on a planet. Using their Equation (10), we computed the restricted Jeans escape parameter as $\Lambda = 27$.  \citet{Fossati:2017A&A...598A..90F} show in their Fig. 4 that $\Lambda$ values below the threshold value of $\Lambda_{T} = 15-35$ correspond to significant atmospheric loss. Consequently, TOI-1173 A $b$ might be losing its atmosphere. However, it is important to highlight that this possibility is only speculative at the moment given the lack of transmission spectroscopy or multi-band transit photometry for \toi. Mass loss measurements are necessary for confirmation, as previously mentioned in Section \ref{sec:intro}.

\subsubsection{Other Causes of the Anomalously Large Radius}
Alternatively, explanations like high-altitude dust or hazes have been suggested to account for the inflated radii of super-Neptunes \citep[e.g.,][]{Wang:2019ApJ...873L...1W, Gao:2020ApJ...890...93G}. However, with our current data, it is challenging to confirm if any of these mechanisms could be responsible for inflating the radius of TOI-1173 A $b$, so that transmission spectroscopy is deemed necessary and encouraged for a more conclusive understanding of the planet's atmospheric physics. 

\subsection{Dynamical Timescales}\label{sec:dyntime}

In the following, we explore noteworthy features of the interactions between TOI-1173 A and its companion, focusing on turbulent friction, magnetic field maintenance, and orbital dynamics. When a perturbation occurs, interactions with the primary star can serve as a damping mechanism, effectively obliterating past effects, particularly if these interactions are on a comparable timescale to the age of the system. This phenomenon can be investigated through the consideration of dynamical timescales derived from the mass of the system components (primary star, secondary star, and the planet), the degree of tidal detachment, and their respective orbital eccentricities. 

\subsubsection{Orbital Eccentricity Evolution}

We considered the timescale on which TOI-1173 A $b$'s orbital eccentricity might have evolved over time, expressed by the ratio $\tau_{e} \sim e/(de/dt)$ following the framework presented by \cite{2022ApJ...926L..17R} -- c.f. Equations. (6) to (13). We measured an eccentricity of $0.023$ for TOI-1173 A $b$, and it is worth noting that, over time, planets are anticipated to circularise their orbits especially due to energy loss caused by tidal dissipation \citep{2018ARA&A..56..175D}. We adopted in this case a tidal dissipation parameter within the range $\mathcal{Q}^\prime = 10^4 - 10^6$ and $\kappa_2 = 0.3$ as the Love number, in-line with \cite{2018AJ....155..165P} and \cite{2022ApJ...926L..17R}. 
Note that a value of $\mathcal{Q}^\prime$ in this range is also supported by the results obtained in Sec. \ref{subsec:excessinternalheat} using the tidal heating model of \citet{Millholland:2020ApJ...897....7M}. 

Based on the physical and orbital parameters of TOI-1173 A $b$, namely orbital period $P_b$, eccentricity $e$, semi-major axis $a_b$ and mass $M_b$ (see Table \ref{tab:allesfitter_results}), we calculated a tidal circularization timescale of $\tau_e \approx 10^{8}-10^{10}$ years for TOI-1173 A $b$.

Relative to the system's age, the tidal circularization timescale could be as short as 100 Myr, which is much less than the age of the planet ($\Gamma \sim 8.7$ Gyr). Therefore, it is plausible that TOI-1173 A $b$ may have exhibited a higher eccentricity in the past, and it was subsequently dampened after a few billion years. On the other hand, the circularization timescale may also be much longer, such that in this case, no evolution in eccentricity over the planet's lifetime would be expected. Since the derived eccentricity of \toi\ today is consistent with zero at the 1.5~$\sigma$ level, we cannot draw strong conclusions about the original eccentricity of its orbit at formation.

\subsubsection{Perturbations due to a Tertiary Companion}

The planet's orbit may even have the potential to adopt a low-eccentricity arrangement through von-Zeipel-Lidov-Kozai (vZLK) cycles due to interactions with a tertiary companion (see \citealt{2023AJ....165...65R}). Adopting a quadrupole level of approximation, and considering that the planet's mass is considerably smaller than the proper masses of the binary stars, we used Equation (27) in \citet{2016ARA&A..54..441N} to estimate the timescale at which the vZLK process may play a part in the evolution of TOI-1173 A $b$. 

Because the eccentricity for the TOI-1173 A/B system is uncertain, to estimate $\tau_{\rm vZLK}$, we considered $e_B$ ranging from $0.0$ to $0.9$. At the current architecture, we estimated the timescale for vZLK perturbations in the order of $\tau_{\rm vZLK, e = 0} \sim 10^{13}$ years, and $\tau_{\rm vZLK, e = 0.9} \sim 10^{12}$ years. Comparing those parameters with the system's age, we observed that the timescale is bigger than $\Gamma$, much longer than the age of the Universe in the null-eccentricity case, indicating that vZLK migration might not have significantly influenced the system's evolution if TOI-1173 A $b$ originated near its present location, i.e., its damping timescale is sufficiently large to be unobserved. 

Interestingly, when extrapolating its initial position for $P_b = 1$ year, a Jupiter-like current location with $P_b \sim 12$ years, or Saturn-like with $P_b \sim 30$ years, $\tau_{\rm vZLK}$ ranges between $\sim 10^{11}$ and $\sim 10^{8}$ years. If the planet initially formed far away from its current position, consistent with theories of giant planet formation (\citealt{2008ApJ...685..560D, 2015PNAS..112.4214B}, and references therein), $\tau_{\rm vZLK}$ approaches $\Gamma$, indicating the possibility of past migration through vZLK mechanisms. 

\subsubsection{Apsidal Precession due to General Relativity}

vZLK oscillations can be however dampened by additional perturbations that induce apsidal precession at a faster rate, thereby diminishing the orbit-averaged torque exerted by the companion star \citep{2003ApJ...589..605W}. As for apsidal precession arising from general relativity, the dampening timescale is given in Equation (29) in \cite{2023AJ....165...65R}. 

For TOI-1173 A $b$ we estimated $\tau_{\rm GR} \sim 10^4$ years; a short timescale when compared to both $\tau_{\rm \tiny vZLK}$ and the system's age. This effectively eliminates the possibility of the system undergoing vZLK oscillations, as the influence of general relativity on apsidal precession is substantially faster when compared to the system's age.

\section{Conclusions and Future Work}
\label{sec:conclusions}

We present the discovery of TOI-1173 A $b$, the first super-Neptune ($M = 27.4 M_\oplus$) with a highly inflated radius ($R = 9.19 R_{\oplus}$) in a wide binary system (projected separation $\sim 11,400$ AU). TESS photometry and radial velocity observations with the MAROON-X and HIRES spectrographs revealed a planet on a nearly circular ($e = 0.023$) close-in orbit ($P = 7.064$ days, $a = 0.069$ AU) to the primary star. Additionally, we infer a density of $\rho = 0.195$ g cm$^{-3}$, equilibrium temperature $T_{\rm{eq}} = 868$ K and insolation flux S$_{\oplus} = 132$ for the planet. TOI-1173 A $b$ stands out as the only low-density super Neptune in a binary system reported to date. 

Hypotheses exploring the causes of the planet's inflated nature include stellar insolation, excess internal heat, presence of planetary rings, and the potential mass loss by atmospheric escape. Although stellar insolation, internal heat and mass loss may contribute (see Subsection \ref{sec:aaa}), we highlighted the negligible impact of planetary rings due to the planet's tidally locked nature. Therefore, we conclude that the most promising scenario to explain the inflated radius of TOI-1173 A $b$ is tidal heating.

We investigated the dynamical evolution of TOI-1173 A and its planetary companion, examining factors such as eccentricity evolution influenced by tidal dissipation, the impact of tertiary companions, and the role of general relativity-induced apsidal precession (see Sec. \ref{sec:dyntime}). 

Whether the system was born aligned or misaligned, however, is uncertain with current data, and TOI-1173 A $b$ represents a benchmark follow-up case for further analysis via the Rossiter-McLaughlin (RM) effect. Following the framework in \cite{2018haex.bookE...2T}, we estimated an expected RM amplitude of $\sim6$ m s$^{-1}$ for TOI-1173 A $b$, so that the system could be observed with current extreme-precision spectrographs such as the Keck Planet Finder (KPF; \citealt{2016SPIE.9908E..70G}),  the EXtreme PREcision Spectrometer (EXPRES; \citealt{2020AJ....159..238B}), and/or the M dwarf Advanced Radial velocity Observer Of Neighboring eXoplanets (MAROON-X; \citealt{2018SPIE10702E..6DS}). 

The relative comparability between the tidal circularization timescale and the system's age, expressed as $\tau_e \sim \Gamma$, leaves us without a clear conclusion on whether the system exhibited a high eccentricity in the past. Under these highly eccentric configurations, other possible formation mechanisms include single planet-planet scattering events (e.g., \citealt{1996Sci...274..954R, 2008ApJ...686..580C}) and resonant interactions \citep{2011Natur.473..187N, 2016ApJ...829..132P}, while low-eccentricities are usually associated with high multiplicity, as in our own Solar System (e.g., \citealt{1993ARA&A..31..129L, 2008ApJ...686..621F, 2015PNAS..112...20L, 2015ApJ...808..126V}) or smooth disc migration \citep{2003ApJ...585.1024G, 2021A&A...646A.166L}. With $\tau_{\rm vZLK} > \Gamma$, TOI-1173 A $b$ might have undergone migration through this mechanism, however, the possibility of current vZLK oscillations occurring in the system is eliminated given that the apsidal precession timescale due to general relativity is orders of magnitude smaller than the proper vZLK timescale ($\tau_{\rm GR} \ll \tau_{\rm vZLK})$.

Looking ahead, the low mass and density of TOI-1173 A $b$, along with TOI-2525 $b$, WASP-107 $b$, TOI-1420 $b$, HATS-8 $b$, and WASP-193 $b$, make it a key planet for atmospheric and dynamical characterization, with further analysis necessary and encouraged to uncover its extreme atmospheric physics in details, as well as to reveal its past orbital configuration. 

\section*{Acknowledgments}

Jhon Yana Galarza acknowledges support from a Carnegie Fellowship. Thiago Ferreira acknowledges support from Yale Graduate School of Arts and Sciences. Diego Lorenzo Oliveira acknowledges  support from CNPq (PCI 301612/2024-2). Henrique Reggiani acknowledges the support from NOIRLab, which is managed by the Association of Universities for Research in Astronomy (AURA) under a cooperative agreement with the National Science Foundation. 

This work made use of data collected with the Gemini Telescope. This research has made use of the Keck Observatory Archive (KOA), which is operated by the W. M. Keck Observatory and the NASA Exoplanet Science Institute (NExScI), under contract with the National Aeronautics and Space Administration. This paper includes data collected by the TESS mission. Funding for the TESS mission is provided by NASA's Science Mission Directorate. This research has made use of the Exoplanet Follow-up Observation Program (ExoFOP; DOI: 10.26134/ExoFOP5) website, which is operated by the California Institute of Technology, under contract with the National Aeronautics and Space Administration under the Exoplanet Exploration Program. Some of the data presented herein were obtained at Keck Observatory, which is a private 501(c)3 non-profit organization operated as a scientific partnership among the California Institute of Technology, the University of California, and the National Aeronautics and Space Administration. The Observatory was made possible by the generous financial support of the W. M. Keck Foundation. The authors wish to recognize and acknowledge the very significant cultural role and reverence that the summit of Maunakea has always had within the Native Hawaiian community. We are most fortunate to have the opportunity to conduct observations from this mountain. Some of the data presented in this paper were obtained from the Mikulski Archive for Space Telescopes (MAST) at the Space Telescope Science Institute. The specific observations analyzed can be accessed via \dataset[http://dx.doi.org/10.17909/dpx3-gv19]{http://dx.doi.org/10.17909/dpx3-gv19}. STScI is operated by the Association of Universities for Research in Astronomy, Inc., under NASA contract NAS5–26555. Support to MAST for these data is provided by the NASA Office of Space Science via grant NAG5–7584 and by other grants and contracts.

\vspace{4mm}
%\clearpage 
\facilities{Gemini, Keck, TESS, The Encyclop{\ae}dia of Exoplanetary Systems, The Exoplanet Follow-up Observing Program.}

\software{
\textsc{numpy} \citep{van_der_Walt:2011CSE....13b..22V}, 
\textsc{matplotlib} \citep{Hunter:4160265}, 
\textsc{pandas} \citep{mckinney-proc-scipy-2010}, 
\textsc{lightkkurve} \citep{2018ascl.soft12013L}, 
\textsc{scipy} \citep{2020NatMe..17..261V},
\textsc{juliet} \citep{2019MNRAS.490.2262E}, 
\textsc{AstroML} \citep{2014sdmm.book.....I}, 
\textsc{DYNESTY} \citep{2020MNRAS.493.3132S}, 
\textsc{allesfitter} \citep{2021ApJS..254...13G}, 
\textsc{astrobase} \citep{2021zndo...4445344B},
\textsc{SERVAL} \citep{Zechmeister:2020ascl.soft06011Z},
\textsc{MESA} \citep{Paxton:2019ApJS..243...10P}, 
\textsc{emcee} \citep{2013PASP..125..306F}, 
\textsc{Sub-Saturns}
\citep{Millholland:2020ApJ...897....7M}, 
\textsc{smplotlib} 
\citep{jiaxuan_li_2023_8126529}.
}

\bibliography{bib}{}
\bibliographystyle{aasjournal}

\end{document}